\pdfoutput=1
\documentclass[11pt,a4paper]{article}
\usepackage{jheppub}
\usepackage[caption=false]{subfig}
\newcommand{\beq}{\begin{eqnarray}}
\newcommand{\eeq}{\end{eqnarray}}
\newcommand{\non}{\nonumber\\}
\newcommand{\p}{\partial}

\def\sech{\qopname\relax o{sech}}

\def\U{\qopname\relax o{U}}

\title{Chemical bonds of two vortex species with a generalized
  Josephson term and arbitrary charges}

\author{Chandrasekhar Chatterjee$^1$,}
\author{Sven Bjarke Gudnason$^2$, and}
\author{Muneto Nitta$^1$}
\affiliation{$^1$Department of Physics, and Research and Education Center
  for Natural Sciences, Keio University, Hiyoshi 4-1-1, Yokohama,
  Kanagawa 223-8521, Japan}
\affiliation{$^2$Institute of Contemporary Mathematics, School of
  Mathematics and Statistics, Henan University, Kaifeng, Henan 475004,
  P.~R.~China} 
\emailAdd{chatterjee(at)keio.jp}
\emailAdd{gudnason(at)henu.edu.cn}
\emailAdd{nitta(at)phys-h.keio.ac.jp}

\abstract{
We consider the Abelian-Higgs model with two complex scalar fields
and arbitrary positive integer charges with the addition of 
a higher-order generalization of the Josephson term.
The theory possesses vortices of both local and global variants.
The only finite-energy configurations are shown to be the local
vortices for which a certain combination of vortex numbers and
electric charges -- called the global vortex number -- vanishes.
The local vortices have rational fractional magnetic flux, as opposed
to the global counterparts that can have an arbitrary fractional
flux.
The global vortices have angular domain walls, which we find good
analytic approximate solutions for.
Finally, we find a full classification of the minimal local vortices
as well as a few nonminimal networks of vortices, using numerical
methods. 
}

\begin{document}
\maketitle
\addtolength{\parskip}{5pt}

\section{Introduction}
A type-II superconductor expels an applied magnetic field due to the 
Meissner effect, such that it flows around the superconducting
material -- with the exception of formation of magnetic vortices,
called Abrikosov-Nielsen-Olesen (ANO) vortices
\cite{Abrikosov:1956sx,Nielsen:1973cs}.
The latter can penetrate the superconductor in the direction of the
externally applied magnetic field if the magnetic field is not too
large and not too small. 
ANO vortices
are elegantly described in the
Ginzburg-Landau (GL) effective theory of superconductivity. 
The static vortices are independent of the theory being relativistic
or nonrelativistic.
$n$ vortices are described by $n$ zeroes in the order parameter of the
theory, which is a complex scalar field, and has additionally a
phase that winds $n$ times at spatial infinity (or at any
radius far from the centers of the $n$ zeroes), see
e.g.~the text book \cite{Jaffe:1980}. 

The order parameter of a superconductor in the GL
effective theory has electric charge $2e$.
In condensed matter physics, liquid metallic deuterium hosts
both a Bose-Einstein condensate (BEC) as well as Cooper pairs
\cite{Babaev:2004rm}.
Such a system can be described by two coupled fields with opposite
signs for their respective electric charges.
Mixtures of condensates of different electric charges is possible also
in ultra-cold atomic gases, with a synthetic gauge field, see
e.g.~ref.~\cite{Zohar:2012ay}.
Two-dimensional Skyrmions induced by quadratic band touching fermions
can provide superconductivity with charge $4e$ instead of $2e$
\cite{Moon:2012jf}.
These facts motivate the study of Abelian-Higgs type models with generic
electric charges. 
Condensed matter systems with two superconductors separated physically
by a thin insulating or weakly superconducting layer are described by
a two-component Abelian Higgs model, see
refs.~\cite{Babaev:2002ck,Herland:2010}.
The case of a two-component superconductor with generic charges has
been studied already by Garaud and Babaev \cite{Garaud:2014laa}. 

This physical setup of two superconductors separated by a thin layer
houses a seminal effect called the Josephson effect, where a
superconducting current is passing through the thin layer and is
proportional to sine of the so-called Josephson phase
\cite{Josephson:1962zz,Josephson:1974uf}. 
The Josephson phase is defined as the \emph{difference} between the
arguments of the order parameters of the two respective
superconductors sandwiching the thin insulating layer.
The effect can be understood quantum mechanically as a tunneling of a
Cooper pair across the thin layer and is described in the
two-component Abelian-Higgs model by the so-called Josephson term. 
Because the effect is really quantum mechanical -- although
macroscopically observable -- it is possible to construct a
single-electron transistor using the Josephson junction
\cite{Fulton:1989}. 
Other interesting applications are superconducting quantum
interference devices called SQUIDs, as well as qubits which are the
integral part of current day's quantum computing, see
ref.~\cite{Wendin:2016} for a review.

The Josephson term also arises in two-component Bose-Einstein
condensates (BECs) as an intrinsic interaction between two different
states of the same condensate, propagated by an external field, as
opposed to the Josephson junction, described above, where the
interaction comes from a tunneling effect between two different
condensates (order parameters). 
The former is called a Rabi term for BECs
\cite{Matthews:1999,Williams:2000,Zibold:2010} and an intrinsic
Josephson term for type-II superconductors. 
Josephson terms have been also studied in field-theoretical setups 
\cite{Nitta:2012xq,Kobayashi:2013ju,Fujimori:2016tmw},
and have been generalized to non-Abelian Josephson junctions
\cite{Nitta:2015mma,Nitta:2015mxa} and intrinsic interactions
\cite{Nitta:2014rxa}.\footnote{The Josephson interaction
  term also appears naturally in two-Higgs doublet models (2HDMs) in
  particle physics (see ref.~\cite{Branco:2011iw} for a review), which
  can affect vortices and give rise to domain walls
  \cite{Eto:2018hhg,Eto:2018tnk}.}

Sine-Gordon solitons often exist in two-component models possessing
Josephson terms and live in Josephson phase function
\cite{Ustinov:1998,Tanaka:2001a}.
They have also been found in one-dimensional type-II superconductors
without a Josephson junction \cite{Tanaka:2001b} and in two-component
BECs \cite{Son:2001td}. 
Vortices in each component are fractionally quantized in the sense
that integer winding numbers of the complex fields at spatial infinity
lead to a fractional magnetic flux \cite{Babaev:2001hv,Babaev:2004rm}.
The presence of Josephson terms induces attraction between the
vortices of the two different components, which creates the
possibility of fractional vortex molecules
\cite{Son:2001td,Kasamatsu:2004tvg}. 
Fractional vortex molecules are subject to vortex confinement 
\cite{Goryo:2007,Tylutki:2016mgy,Eto:2017rfr} and can give rise to
vortex lattices \cite{Cipriani:2013nya} as well as
Berezinskii-Kosterlitz-Thouless transitions \cite{Kobayashi:2018ezm}.

In this paper, we consider the marriage of the two concepts: namely the
possibility of having generic charges of the two order parameters in a
two-flavor model (or two-component superconductor) and having a
Josephson term.
This has not been considered previously, for good reasons, because the
Josephson term only allows the two order parameters to have the same
charges.
In this paper we propose a generalization of the Josephson term to
arbitrary integer charges of the same sign, which is gauge invariant,
but does not have the physical interpretation of a Cooper pair
tunneling across a thin barrier, unless the two charges are both equal
to $2$ (from now on, we will only specify the integer coefficient of
the charge $e$).

Our first result is that the total winding number $k$, is a rational
fraction when the vortices are local -- i.e.~when $n_{\rm global}=0$,
but is arbitrarily fractional (could be irrational) when the vortices
are global -- i.e.~when $n_{\rm global}\neq 0$.
The latter situation yields angular domain walls, which we find
analytic approximations for.
The angular domain walls live in the generalized Josephson phase,
which we denote the global vortex phase.

Next, we find that the only finite-energy vortices that exist in our 
model are local vortices -- i.e.~with
$n_{\rm global}=0$.
If the vortices are global, on the other hand, the energy is
logarithmically divergent (in the plane) without the new generalized
Josephson term and linearly divergent with it.
The latter linear divergence is simply due to the presence of the
angular domain walls emanating from the vortex cores of the global
vortices.

Finally, we perform numerical calculations and find a full
classification of the minimal local vortices in the theory for
electric charges up to four.
We also illustrate some nonminimal local vortices that come in two
variants of infinitely extendable networks.
Finally, we make a numerical comparison between the angular domain
walls and the analytic solution.

The paper is organized as follows.
Sec.~\ref{sec:model} introduces the model, the symmetries in
sec.~\ref{sec:symmetries} and the vacuum in sec.~\ref{sec:vac}.
The vortex Ansatz is used to construct local and global vortices in
sec.~\ref{sec:vortex_ansatz} with and without the generalized
Josephson term and in the former case, an approximation makes it
possible to find an explicit analytic solution for the so-called
angular domain wall, that exists for global vortices.
Finite energy considerations are made in sec.~\ref{sec:finite_energy}. 
The results are presented in sec.~\ref{sec:results}, first describing
the numerical method (sec.~\ref{sec:num_method}), the visualization
scheme (sec.~\ref{sec:vis}) and some intuition for constructing
solutions (sec.~\ref{sec:intuitive}).
Then local vortices are constructed in
sec.~\ref{sec:constructing_local}, whose minimal solutions are
illustrated in sec.~\ref{sec:minimal_local} and nonminimal in
sec.~\ref{sec:nonminimal_local}.
The dependence of the ratio of the VEVs is explored in
sec.~\ref{sec:varying_vevs}.
Global vortices are considered in sec.~\ref{sec:global}, where a
numerical solution is compared with the analytic approximate solution
of sec.~\ref{sec:vortex_ansatz}.
Finally, the paper is concluded with a discussion in
sec.~\ref{sec:conclusion}.

\section{The model, vortex asymptotics and angular domain walls}\label{sec:model}

The static energy for our model containing two complex scalar
fields (two components) $\phi_1$ and $\phi_2$ with charges $Q_1$ and
$Q_2$, respectively, reads 
\begin{align}
  E &= \int d^2x\left[
  \frac{1}{4e^2}F_{ij}^2
  +|D_i\phi_f|^2
  +\sum_{f=1}^2\frac{\lambda_f^2}{2}\left(|\phi_f|^2 - v_f^2\right)^2
  -\gamma\left(\phi_1^{Q_2}\bar{\phi}_2^{Q_1} + \bar{\phi}_1^{Q_2}\phi_2^{Q_1}\right)
  +\eta\right],
  \label{eq:energy}
\end{align}
and the covariant derivatives are given by
\beq
D_i\phi_f = \p_i\phi_f - iQ_f A_i\phi_f, \qquad \textrm{($f$ not summed over)}
\eeq
with the spatial index $i$ running over $1,2$
and the flavor index $f=1,2$.
The coupling constants of the model are $e>0$ for the $\U(1)$ gauge
coupling constant, $Q_f\in\mathbb{Z}_{>0}$ are positive integer
charges, $\lambda_f>0$ for the two symmetry breaking
potentials, $v_f>0$ for the vacuum expectation values of the norm of
the complex scalar fields, and finally $\gamma$ for the generalized
Josephson term.
$\eta\in\mathbb{R}$ is real constant that will be adjusted so as to
set the vacuum energy to zero.
One reason for considering only positive charges
$Q_f\in\mathbb{Z}_{>0}$ is that the generalized Josephson term is then
well defined also at the points in space where
$\sum_{g=1}^2\sigma^1_{f g}\phi_g$ vanishes, with
\beq
\sigma^1_{f g} =
\begin{pmatrix}
  0 & 1\\
  1 & 0
\end{pmatrix}_{f g},
\eeq
being the first Pauli matrix and $f,g$ are the matrix indices.

The equations of motion read
\begin{align}
  D_i^2\phi_f
  -\lambda_f^2\left(|\phi_f|^2 - v_f^2\right)\phi_f
  +\gamma \sum_{g=1}^2\sigma^1_{f g}Q_g\phi_g^{Q_f}\bar{\phi}_f^{Q_g-1} &= 0, \qquad \textrm{($f$ not summed over)}\label{eq:eom_phi}\\
  \p_i F_{ij}
  - i e^2\sum_{f=1}^2 Q_f \left(\bar{\phi}_f D_j\phi_f - \phi_f \overline{D_j\phi}_f\right) &= 0.\label{eq:eom_A}
\end{align}

The generalized Josephson term will lead to a potential runaway
instability (i.e.~unbounded energy functional) for $Q_f>4$ or for
$Q_f=4$ and $\gamma>\lambda_f^2$.
For this reason, we will not consider charges larger than $Q_f=4$ for
any $f$.

We note that the sign of the coefficient of the generalized Josephson
term, $\gamma$, is unphysical. This can be seen readily by performing
a global transformation of either of the complex scalar fields:
e.g.~take $\phi_1$ and make the transformation
$\phi_1\to e^{i\frac{\pi}{Q_2}}\phi_1$, which changes
$\gamma\to-\gamma$.

\subsection{Symmetries}\label{sec:symmetries}

The symmetry of the model for $\gamma=0$ is
\beq
G = \U(1)_1\times\U(1)_2,
\eeq
which is a global symmetry
\begin{align}
\U(1)_f\ : \quad \phi_f \to e^{i\beta_f}\phi_f, \qquad\textrm{($f$ not summed over)}
\end{align}
with $f=1,2$.
Turning on $\gamma>0$ breaks the symmetry down to a subgroup -- a
combination of the two $\U(1)$'s:
\beq
\phi_1 \to e^{iQ_1\beta}\phi_1, \qquad
\phi_2 \to e^{iQ_2\beta}\phi_2.
\eeq
This is the combination that is gauged and thus the gauge
transformation reads
\beq
\phi_1 \to e^{iQ_1\alpha(x)}\phi_1, \qquad
\phi_2 \to e^{iQ_2\alpha(x)}\phi_2, \qquad
A_i \to A_i + \p_i\alpha(x).
\eeq
The model is invariant under this $\U(1)$ gauge transformation;
in particular the generalized Josephson term is constructed on the
principle of \emph{gauge invariance}.

The breaking of the symmetry $G\to\tilde{G}$ is explicit (for
$\gamma>0$) with
\beq
\tilde{G} = \U(1),
\eeq
and this symmetry is then further completely broken spontaneously 
\beq
\tilde{G} \to H, \qquad \textrm{with} \qquad
H = \mathbf{1}.
\eeq
Thus the topology of the vacuum supports topological vortices
\beq
\pi_1(\tilde{G}/H) = \pi_1(\U(1)) = \pi_1(S^1) = \mathbb{Z}.
\eeq

Although the symmetry $G$ is explicitly broken, it will be useful to
consider the generalized Josephson term as a perturbation and hence
remember the original symmetry of the model, $G$, which gives rise to
two distinct vorticities or vortex numbers
\beq
\pi_1(G/H) = \pi_1\left(\U(1)_1\times\U(1)_2\right)
= \mathbb{Z}\oplus\mathbb{Z}.
\eeq
If we denote the two vortex numbers by $n_f$:
\beq
n_f \in \mathbb{Z}, \qquad f=1,2,
\eeq
it will prove convenient to define two linear combinations
\begin{align}
  n_{\rm local} &= Q_2 n_1 + Q_1 n_2, \label{eq:nlocal}\\
  n_{\rm global} &= Q_2 n_1 - Q_1 n_2. \label{eq:nglobal}
\end{align}
The meaning of these new ``vortex numbers'' will become clear
shortly; they can be viewed as the number of local and number of
global vortices, respectively.

\subsection{The vacuum}\label{sec:vac}

Minimization of the Mexican hat potential (the third term in the
energy \eqref{eq:energy}) leads to the vacuum solution
\beq
\phi_f = v_f e^{i\theta_f}. \qquad \textrm{($f$ not summed over)}
\eeq
Substituting this into the generalized Josephson term (the fourth term
in the energy \eqref{eq:energy}) gives
\beq
-2\gamma v_1^{Q_2}v_2^{Q_1}
\cos(Q_2\theta_1-Q_1\theta_2).
\eeq
The number of vacua is infinite, because we can shift both angles and
get a ``new vacuum'': we will count such a vacuum as the same.
Counting how many discrete possibilities the problem 
\beq
Q_2\theta_1 - Q_1\theta_2 = \frac{\pi}{2} + \mathsf{m}\pi,
\label{eq:theta_vac_solns}
\eeq
with $\mathsf{m}\in\mathbb{Z}$ has for $\theta_{1,2}\in[0,2\pi)$
yields the result 
\beq
\#\textrm{ of vacua} = 2\max\left(Q_1,Q_2\right),
\eeq
where we recall that the charges are positive integers:
$Q_f\in\mathbb{Z}_{>0}$.
This means that the number of \emph{different} domain walls is
\beq
2\max\left(Q_1,Q_2\right) - 1.
\eeq

The constant $\eta$ is thus determined as
\beq
\eta = 2\gamma v_1^{Q_2}v_2^{Q_1}
\cos(Q_2\theta_1-Q_1\theta_2),
\eeq
which vanishes ($\eta=0$) for the $2\max(Q_1,Q_2)$ solutions
\eqref{eq:theta_vac_solns}, but otherwise is generically equal to
\beq
\eta = 2\gamma v_1^{Q_2}v_2^{Q_1}.
\eeq

\subsection{Vortex Ansatz}\label{sec:vortex_ansatz}

\subsubsection{\texorpdfstring{$\gamma=0$}{gamma=0}}

We consider the following axially symmetric Ansatz for vortices in
both complex scalar fields and a corresponding transverse component of
the gauge field.
\begin{align}
\phi_f &= v_f h_f(r) e^{i n_f\theta}, \qquad \textrm{($f$ not summed over)}\label{eq:axial_hf}\\
A_i &= -\frac{\varepsilon_{ij}x^j}{r^2}k a(r),\label{eq:axial_a}
\end{align}
with boundary conditions
\begin{align}
  h_{1,2}(0) &= 0, &\qquad
  \lim_{r\to\infty}h_{1,2}(r)&=1, \label{eq:axial_BCh}\\
  a(0) &= 0, &\qquad
  \lim_{r\to\infty}a(r)&=1.
  \label{eq:axial_BC}
\end{align}
Writing now the equations of motion
\eqref{eq:eom_phi}-\eqref{eq:eom_A} using the above Ansatz, we have 
\begin{align}
h_f''
+\frac{1}{r}h_f'
-\frac{1}{r^2}(k a Q_f - n_f)^2 h_f
-\lambda_f^2 v_f^2(h_f^2 - 1) h_f &\label{eq:EOMhf}\\
\mathop+\gamma\sum_{g=1}^2\sigma^1_{f g}Q_g\cos\left[(Q_2n_1-Q_1n_2)\theta\right]
  v_f^{Q_g-2}v_g^{Q_f}h_f^{Q_g-1}h_g^{Q_f} &= 0,
  \qquad\textrm{($f$ not summed over)}\non
  \gamma\sum_{g=1}^2\epsilon_{f g} Q_g \sin\left[(Q_2 n_1 - Q_1 n_2)\theta\right]
  v_f^{Q_g-2} v_g^{Q_f} h_f^{Q_g-2} h_g  &= 0,
  \qquad\textrm{($f$ not summed over)}\label{eq:EOMThetaf}\\
k^2\left(a'' - \frac1r a'\right)
-2e^2\sum_{f=1}^2v_f^2h_f^2\left(k a Q_f - n_f\right) k Q_f &= 0,
\label{eq:EOMa}
\end{align}
where $h'\equiv\p_rh$, $h''\equiv\p_r^2h$ and so on.
Unless $Q_2n_1-Q_1n_2=n_{\rm global}=0$, the generalized Josephson
term is not compatible with the axially symmetric vortex Ansatz and we
have to set $\gamma=0$ due to eq.~\eqref{eq:EOMThetaf}, which reduces
eq.~\eqref{eq:EOMhf} to:
\beq
h_f''
+\frac{1}{r}h_f'
-\frac{1}{r^2}(k a Q_f - n_f)^2 h_f
-\lambda_f^2 v_f^2(h_f^2 - 1) h_f = 0.
\qquad\textrm{($f$ not summed over)}
\label{eq:EOMhf2}
\eeq

Applying the boundary conditions \eqref{eq:axial_BC} to
eq.~\eqref{eq:EOMa} yields an algebraic equation that can be solved
for the total winding number, $k$, giving 
\beq
k = \frac{n_1Q_1v_1^2 + n_2Q_2v_2^2}{Q_1^2v_1^2 + Q_2^2v_2^2}.
\label{eq:k}
\eeq
The total magnetic flux is thus
\beq
\Phi
= \int d^2x\; F_{12}
= 2\pi k.
\label{eq:flux}
\eeq
We will call a vortex configuration with this magnetic flux a
\emph{fractional vortex}, because $k$ is generically not an integer
and not even necessarily a rational number.
Rewriting the total winding number in terms of the local and global
vortex numbers \eqref{eq:nlocal}-\eqref{eq:nglobal}, we get
\beq
k = \frac{1}{2Q_1Q_2}n_{\rm local}
 +\frac{Q_1^2v_1^2 - Q_2^2 v_2^2}{2Q_1Q_2\left(Q_1^2v_1^2+Q_2^2v_2^2\right)}n_{\rm global},
\eeq
from which it is clear that $n_{\rm global}$ creates an arbitrary
fraction, whilst $n_{\rm local}$ yields a rational number.

Considering now the asymptotic behavior of the complex scalar fields,
we apply the boundary conditions \eqref{eq:axial_BC} to
eq.~\eqref{eq:EOMhf2} and obtain
\begin{align}
-\frac{1}{r^2}\sum_{g=1}^2\sigma^1_{f g}\left(\frac{n_{\rm global}Q_g v_g^2}{Q_1^2v_1^2+Q_2^2v_2^2}\right)^2.
\label{eq:EOMhf_asymp}
\end{align}
This term comes from the mismatch of the gauge field asymptotically
trying to cancel the contribution from the winding of 
the complex scalar fields.
In fact, the presence of this term (i.e.~for $n_{\rm global}\neq 0$)
will lead to a logarithmic divergence of the energy, see below.

\subsubsection{\texorpdfstring{$\gamma\neq 0$ : Angular domain
    walls}{gamma <> 0 : Angular domain walls}}

We will now consider the following axially symmetric vortex Ansatz,
but with a nontrivial phase function and $\theta$ dependence:
\begin{align}
\phi_f &= v_f h_f(r) e^{i n_f\Theta_f(r,\theta)}, \qquad \textrm{($f$ not summed over)}\label{eq:axial_hf_gamma}\\
A_i &= -\frac{\varepsilon_{ij}x^j}{r^2}k a(r,\theta),\label{eq:axial_a_gamma}
\end{align}
where the boundary conditions now in addition to
eq.~\eqref{eq:axial_BCh} read 
\begin{align}
a(0,\theta) &= 0, &\qquad
\lim_{r\to\infty} a(r,\theta) &= 1, \\
\Theta_{1,2}(r,0) &= 0, &\qquad
\Theta_{1,2}(r,2\pi) &= 2\pi.
\end{align}
Writing out the equations of motion now for the Ansatz
\eqref{eq:axial_hf_gamma}-\eqref{eq:axial_a_gamma} yields
\begin{align}
h_f''
+\frac{1}{r}h_f'
-\frac{1}{r^2}\left(k a Q_f - n_f \dot{\Theta}_F\right)^2 h_f
-n_f^2 h_f (\Theta_f')^2
-\lambda_f^2 v_f^2(h_f^2 - 1) h_f &\label{eq:EOMhf_gamma}\\
\mathop+\gamma\sum_{g=1}^2\sigma^1_{f g}Q_g\cos\left(Q_2n_1\Theta_1-Q_1n_2\Theta_2\right)
  v_f^{Q_g-2}v_g^{Q_f}h_f^{Q_g-1}h_g^{Q_f} &= 0,\non
  \textrm{($f$ not summed over)}\non
n_f \Theta_f''
+\frac{1}{r}n_f\Theta_f'
+2n_f(\log h_f)'\Theta_f'
-\frac{1}{r^2} k Q_f \dot{a}
+\frac{1}{r^2} n_f \ddot{\Theta}_f &\label{eq:EOMThetaf_gamma}\\
-\gamma\sum_{g=1}^2\epsilon_{f g} Q_g v_f^{Q_g-2} v_g^{Q_f}
  h_f^{Q_g-2} h_g \sin(Q_2 n_1 \Theta_1 - Q_1 n_2 \Theta_2) &= 0, \non
  \textrm{($f$ not summed over)}\non
  k \left(a'' - \frac{a'}{r}\right)
  -2e^2 \sum_{f=1}^2Q_fv_f^2h_f^2\left(k a Q_f - n_f \dot{\Theta}_f\right) &= 0,
\label{eq:EOMa_gamma}
\end{align}
where $\epsilon_{fg}$ is the two-dimensional antisymmetric tensor with
$\epsilon_{12}=1$, 
$\Theta'\equiv\p_r\Theta$, $\dot{\Theta}\equiv\p_\theta\Theta$ and so on.
Since we are interested in the asymptotic behavior, we will consider
the approximation of ignoring the radial derivatives.
Suppressing the radial derivatives in eq.~\eqref{eq:EOMa_gamma} and
assuming that $h_f=1$, we get
\beq
k a(\theta) = \frac{Q_1 v_1^2 n_1 \dot{\Theta}_1(\theta) + Q_2 v_2^2 n_2 \dot{\Theta}_2(\theta)}{Q_1^2 v_1^2 + Q_2^2 v_2^2},
\label{eq:asymptotic_a}
\eeq
which reduces to eq.~\eqref{eq:k} if
\beq
\frac{1}{2\pi}\int_0^{2\pi}d\theta\; a = 1.
\eeq
Inserting eq.~\eqref{eq:asymptotic_a} into
eq.~\eqref{eq:EOMThetaf_gamma}, suppressing the radial derivatives and
assuming $h_f=1$, yields 
\begin{align}
  \frac{Q_2 n_1 \ddot{\Theta}_1 - Q_1 n_2 \ddot{\Theta}_2}
       {r^2\left(Q_1^2 v_1^2 + Q_2^2 v_2^2\right)}
- \gamma v_1^{Q_2-2}v_2^{Q_1-2}
  \sin\left(Q_2 n_1 \Theta_1 - Q_1 n_2 \Theta_2\right) = 0.
\label{eq:Thetas1}
\end{align}  
It will prove convenient to switch variables to
\begin{align}
  n_{\rm local}\Theta_{\rm local} &= Q_2 n_1 \Theta_1 + Q_1 n_2 \Theta_2,\\
  n_{\rm global}\Theta_{\rm global} &= Q_2 n_1 \Theta_1 - Q_1 n_2 \Theta_2,
\end{align}
for which eq.~\eqref{eq:Thetas1} becomes
\beq
n_{\rm global} \ddot{\Theta}_{\rm global}
- r^2 \gamma \left(Q_1^2 v_1^2 + Q_2^2 v_2^2\right) v_1^{Q_2-2}v_2^{Q_1-2}
\sin\left(n_{\rm global} \Theta_{\rm global}\right) = 0.
\label{eq:Theta_global}
\eeq
For convenience, we will define the variable
\beq
\zeta \equiv
\theta r \kappa \equiv
\theta r\sqrt{\gamma\left(Q_1^2 v_1^2 + Q_2^2 v_2^2\right)
  v_1^{Q_2-2}v_2^{Q_1-2}}, \label{eq:zeta_kappa_def}
\eeq
in terms of which eq.~\eqref{eq:Theta_global} reads
\beq
n_{\rm global}\frac{d^2\Theta_{\rm global}}{d\zeta^2}
- \sin(n_{\rm global} \Theta_{\rm global}) = 0.
\eeq
This equation is the well-known sine-Gordon equation and the solutions
to this equation are sine-Gordon domain walls -- which we shall call
angular domain walls\footnote{For $Q_1=Q_2=2$
which is the conventional superconductor, the angular domain wall is
called a fluxon \cite{Ustinov:1998}. }
\begin{align}
  \Theta_{\rm global} &= \sum_{\mathsf{m}=0}^{n_{\rm global}-1}
  \Theta_H\left(\theta-\frac{2\pi\mathsf{m}}{n_{\rm global}}\right)
  \Theta_H\left(\frac{2\pi(\mathsf{m}+1)}{n_{\rm global}}-\theta\right)
  \Theta_{\rm global}^{\mathsf{m}}, \label{eq:angularDW}\\ \Theta_{\rm
    global}^{\mathsf{m}} &= \frac{4}{n_{\rm global}}
  \arctan\left(\exp\left[(\theta-\vartheta_{\mathsf{m}})r\kappa\right]\right)
  +\frac{2\pi\mathsf{m}}{n_{\rm global}}, \non \vartheta_{\mathsf{m}}
  &\in \left(\frac{2\pi\mathsf{m}}{n_{\rm global}},
  \frac{2\pi(\mathsf{m}+1)}{n_{\rm global}}\right), \qquad
  \mathsf{m} = 0,1,2,\ldots,n_{\rm global}-1,
\end{align}
where $\Theta_H$ is the Heaviside step function
\beq
\Theta_H(x) = \int_{-\infty}^x ds\; \delta(s),
\label{eq:heaviside_stepfunction}
\eeq
$\vartheta_m$ are directional moduli and $\kappa$ is given by
eq.~\eqref{eq:zeta_kappa_def}.

A comment is in store about the approximation we have made by picking
up only 3 terms of the imaginary part of the scalar field equation of
motion \eqref{eq:EOMThetaf_gamma}.
More precisely, we have used the two terms involving $\theta$
derivatives as well as the potential bit, but discarded the radial
derivatives.
The rationale for this, is that we expect the angular domain walls to
be translational invariant in say $x$, which means that the angular
function will only depend on $r^2$ via the prefactor of the double
$\theta$ derivative.
We check this statement a posteriori, and indeed the radial derivative
of the solution \eqref{eq:angularDW} is exponentially suppressed at
large $r$ and can thus safely be ignored in the asymptotic
determination of the domain wall profile.
It does, however, play a role for the domain wall shape near the
vortex core and this region will also fix the apparent
\emph{directional moduli} of the angular domain walls.
However, to leading order in the asymptotic expansion, we cannot
determine such fine details.

\begin{figure}[!htp]
  \begin{center}
    \mbox{\subfloat[$n_{\rm global}=1$]{\includegraphics[width=0.49\linewidth]{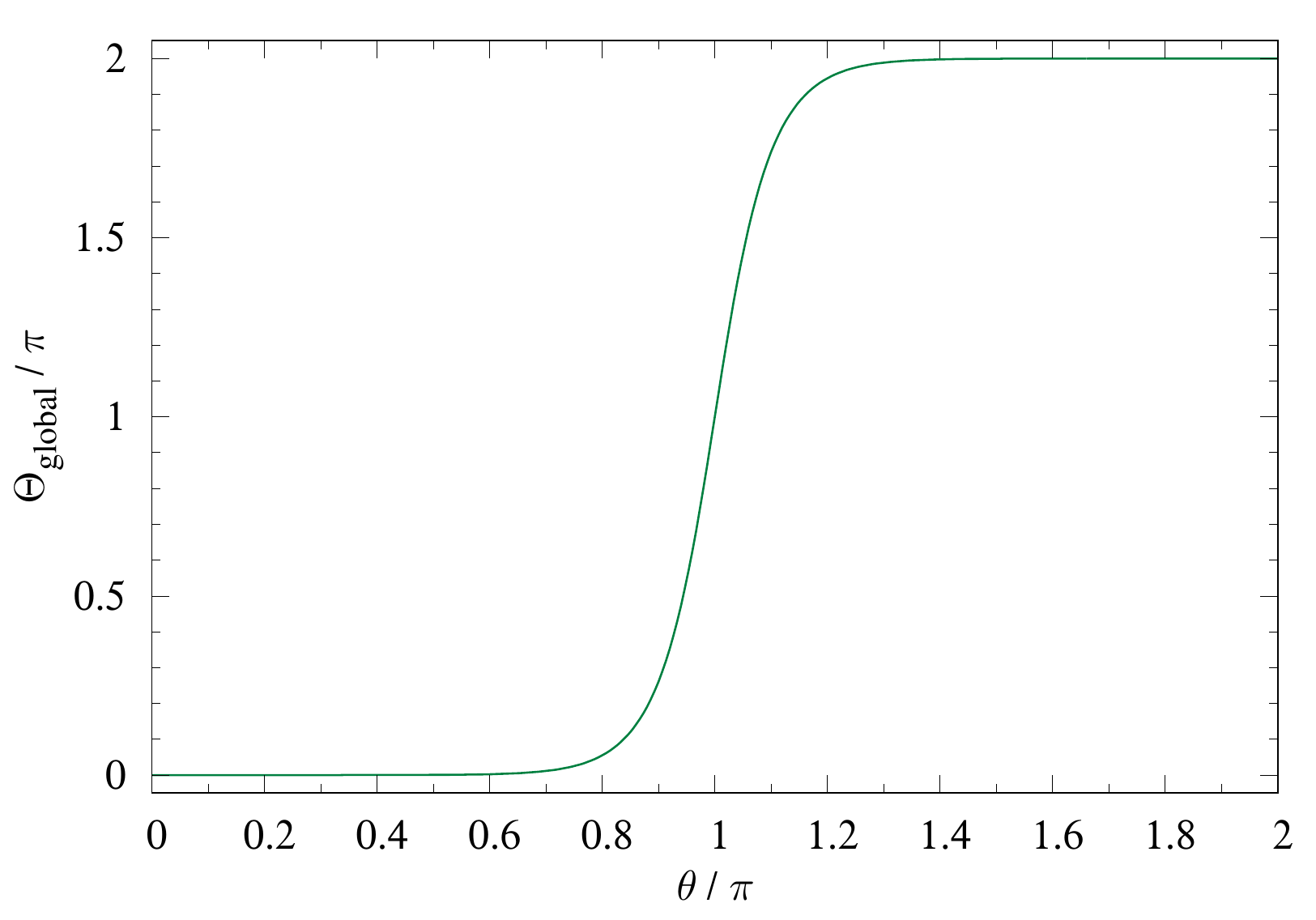}}
      \subfloat[$n_{\rm global}=2$]{\includegraphics[width=0.49\linewidth]{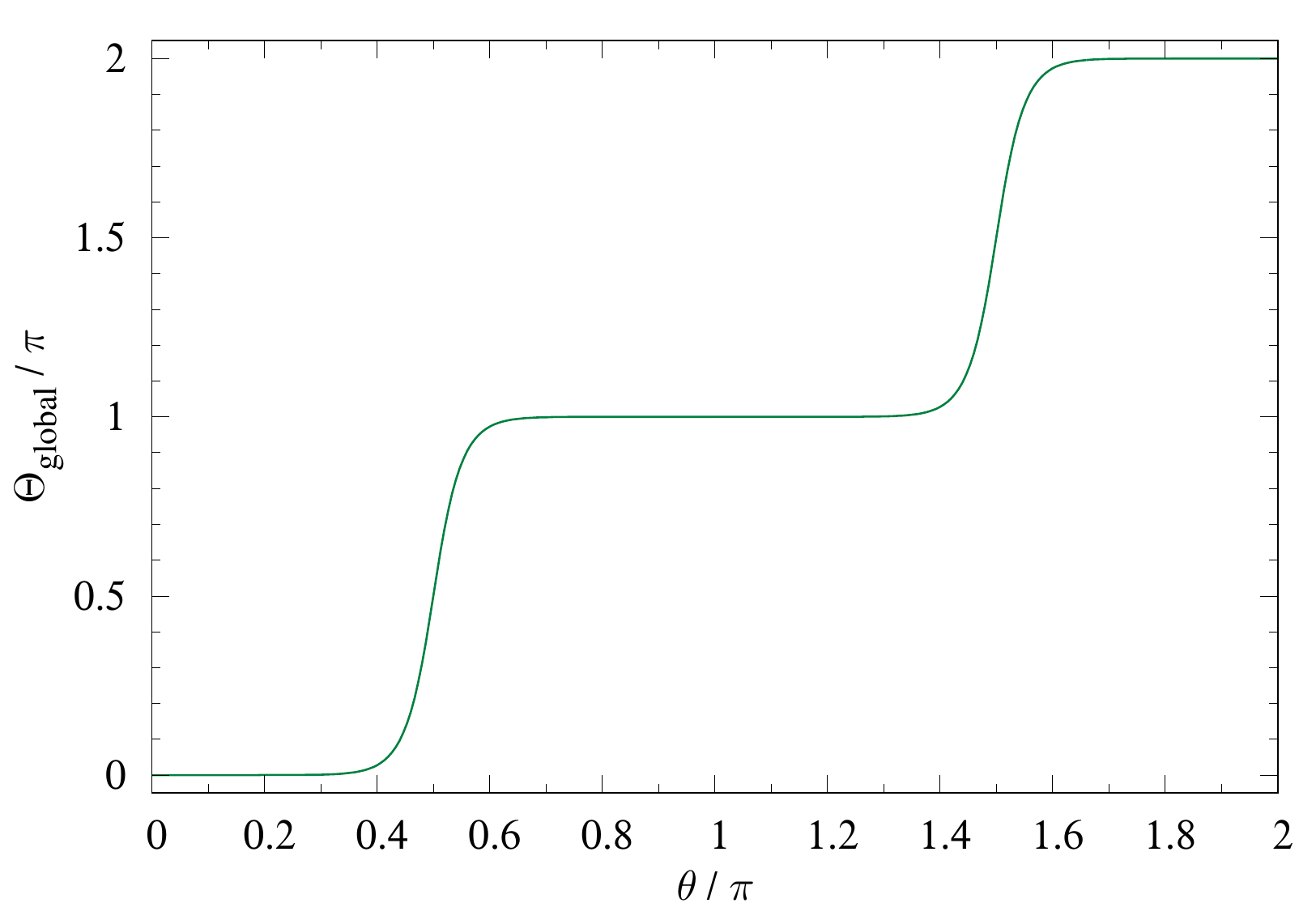}}}
    \mbox{\subfloat[$n_{\rm global}=3$]{\includegraphics[width=0.49\linewidth]{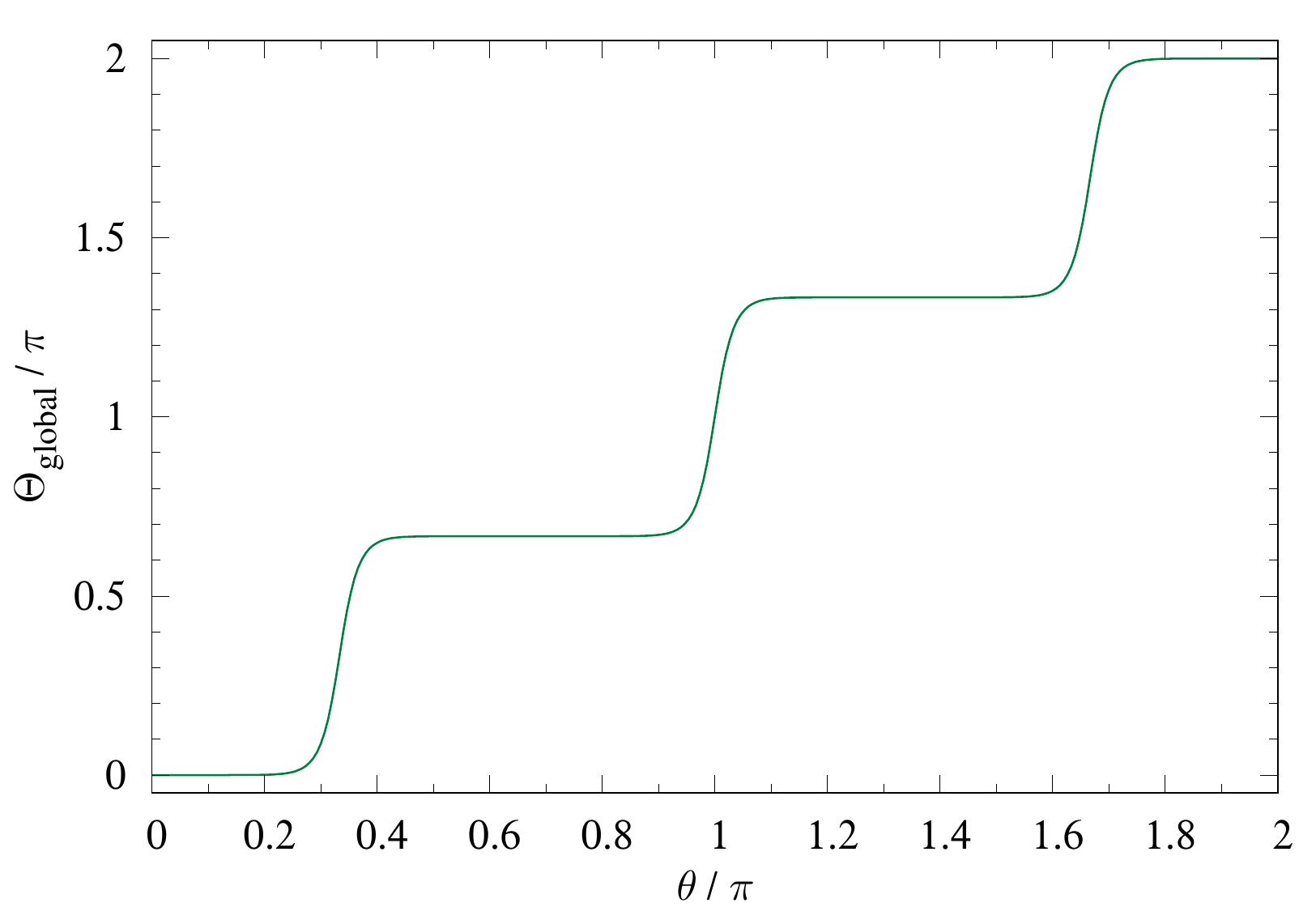}}
      \subfloat[$n_{\rm global}=4$]{\includegraphics[width=0.49\linewidth]{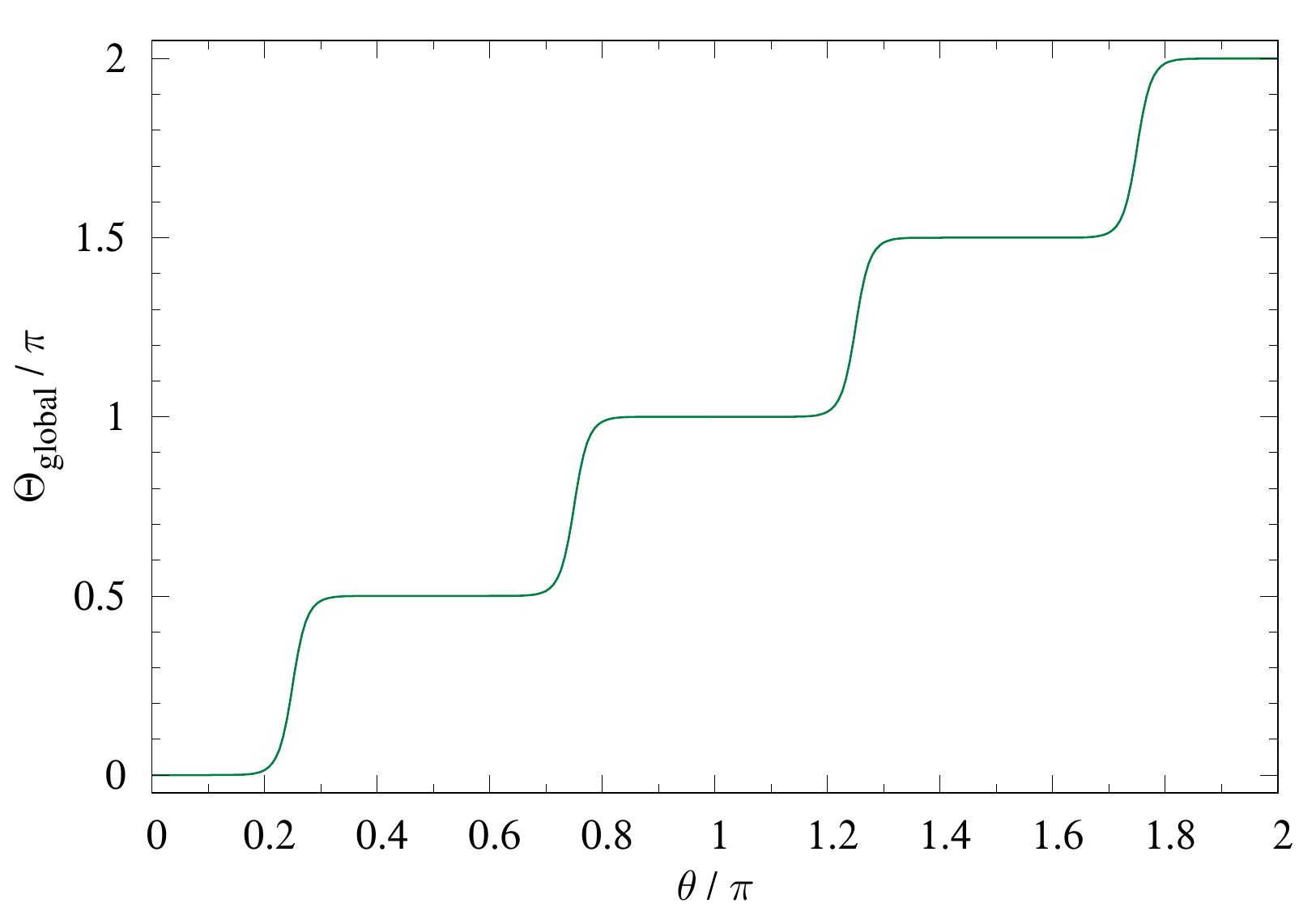}}}
    \caption{Angular domain walls for $n_{\rm global}=1,2,3,4$.
    For the figures, the following values where used:
    $r\kappa=5,10,15,20$ for the four panels, respectively, and the
    directional moduli were set to
    $\vartheta_{\mathsf{m}}=\pi(2\mathsf{m}+1)/n_{\rm global}$,
    with $\mathsf{m}=0,1,\ldots,n_{\rm global}$, which corresponds to
    the middle of their respective intervals (maximal repulsion). }
    \label{fig:adw}
  \end{center}
\end{figure}

Fig.~\ref{fig:adw} shows examples of the angular domain wall solutions
given in eq.~\eqref{eq:angularDW}.
The solution is only valid asymptotically, i.e.~for
$r\kappa\gg n_{\rm global}$. 

Angular domain walls resemble axion domain walls emanating from an
axion string in axion models (for a review, see
e.g.~ref.~\cite{Kawasaki:2013ae}), and axial domain walls from an
axial vortex in QCD (for a review, see e.g.~ref.~\cite{Eto:2013hoa}).

\subsection{Finite energy vortices}\label{sec:finite_energy}

We will now consider finiteness of the total energy in the plane. 
The exact behavior of the vortices near the origin and hence near
their cores will not be important here, merely the behavior of the
fields as they tend to spatial infinity.

\subsubsection{\texorpdfstring{$\gamma=0$}{gamma = 0}}

We start with the case of $\gamma=0$, for which the divergence was
already spotted in eq.~\eqref{eq:EOMhf_asymp}.
For this analysis we again use the Ansatz
\eqref{eq:axial_hf}-\eqref{eq:axial_a} -- which is appropriate for the
$\gamma=0$ case -- and insert it into the energy functional
\eqref{eq:energy}. 
Taking the asymptotic contributions to the energy integral one-by-one,
yields
\begin{align}
  \frac{\pi}{2e^2}\int^R dr\; rF_{ij}^2 &\sim \textrm{finite},\\
  2\pi\int^R dr\; r|D_i\phi_f|^2 &\sim
    \frac{n_{\rm global}^2v_1^2v_2^2}{(Q_1^2v_1^2 + Q_2^2v_2^2)}2\pi\log R,\\
  \pi\sum_{f=1}^2\lambda_f^2\int^R dr\; r\left(|\phi_f|^2 - v_f^2\right)^2 &\sim
    \textrm{finite},
\end{align}
with $R$ much bigger than any other scale in the system.
It is clear that if we impose the condition that
\beq
n_{\rm global} = 0, \qquad \Rightarrow \qquad
n_2 Q_1 = n_1 Q_2,
\eeq
then the total energy is finite, even for infinitely large systems.
Our first result of this paper is, however, this.
Once we impose the condition, $n_{\rm global}=0$ or equivalently
$n_2Q_1=n_1Q_2$, the total winding number of the gauge field
simplifies to
\beq
k = \frac{1}{2Q_1Q_2}n_{\rm local}
= \frac{n_1}{Q_1} = \frac{n_2}{Q_2},
\label{eq:rational_k}
\eeq
and is related to the magnetic flux by eq.~\eqref{eq:flux}.
This winding number corresponds to a finite-energy configuration --
we call a vortex with such a winding number a
\emph{rational (fractional)} vortex.

Notice that the finite-energy configurations do not necessarily have
an integer times $2\pi$ for the flux, but can take on any fractional
value. 
Notice also that for any finite-energy configuration, the values of
$v_{1,2}$ do not affect the total magnetic flux.
For infinite-energy configurations -- global vortices that have
$n_{\rm global}\neq 0$ -- the ratio of $v_1/v_2$ can nevertheless
affect the total magnetic flux unless $|Q_1v_1|=|Q_2v_2|$, for which
an accidental cancellation makes gauge field unaware of the global
vortex winding in the scalar fields. 

\subsubsection{\texorpdfstring{$\gamma\neq 0$ : Angular domain
    walls}{gamma <> 0 : Angular domain walls}}\label{sec:adw}

Turning to the case of $\gamma\neq 0$, axial symmetry of the vortices
is lost and the phase functions of the scalar fields, $\Theta_{1,2}$
-- see eq.~\eqref{eq:axial_hf_gamma}, develop nontrivial behavior
described by angular domain walls \eqref{eq:angularDW}.
For this analysis, we will insert the Ansatz
\eqref{eq:axial_hf_gamma}-\eqref{eq:axial_a_gamma} into the energy
functional \eqref{eq:energy}.
Taking again the asymptotic contributions to the energy one-by-one,
starting with the kinetic terms, yields
\begin{align}
  \frac{\pi}{2e^2}\int^R dr\; rF_{ij}^2 &\sim \textrm{finite},\\
  \int^R dr\; r\int_0^{2\pi}d\theta\; |D_i\phi_f|^2 &\sim
  \int^R dr\; \int_0^{2\pi}d\theta\;
  \frac{1}{r}\sum_{f=1}^2v_f^2
  \left(k a Q_f - n_f \dot{\Theta}_f\right)^2 h_f^2 + \textrm{finite}\non
  &\sim \int^R dr\; \int_0^{2\pi}d\theta\;
  \frac{v_1^2v_2^2n_{\rm global}^2\dot{\Theta}_{\rm global}^2}
       {r(Q_1^2v_1^2 + Q_2^2v_2^2)} \non
  &\sim 4\gamma v_1^{Q_2}v_2^{Q_1}
  \int^R dr\; \int_0^{\frac{2\pi}{n_{\rm global}}}d\theta\; r
  \sum_{\mathsf{m}=0}^{n_{\rm global}-1}
  \sech^2\left[(\theta-\tilde{\vartheta}_{\mathsf{m}})r \kappa\right]\non
  &\sim \frac{4\sqrt{\gamma}v_1^{\frac{Q_2}{2}+1}v_2^{\frac{Q_1}{2}+1}}
  {\sqrt{Q_1^2v_1^2 + Q_2^2v_2^2}}
  \int^R dr \!\!\!\sum_{\mathsf{m}=0}^{n_{\rm global}-1} \!\!\!
  \left[\tanh\left((\theta -\tilde{\vartheta}_{\mathsf{m}})r\kappa\right)\right]_0^{\frac{2\pi}{n_{\rm global}}}\non
  &\sim
  \frac{8\sqrt{\gamma}v_1^{\frac{Q_2}{2}+1}v_2^{\frac{Q_1}{2}+1}n_{\rm
    global}}
  {\sqrt{Q_1^2v_1^2 + Q_2^2v_2^2}} R,
\end{align}
where we have defined the shifted directional moduli
\beq
\tilde{\vartheta}_{\mathsf{m}} \equiv \vartheta_{\mathsf{m}}
- \frac{2\pi\mathsf{m}}{n_{\rm global}}.
\eeq
Turning now to the potentials, we have
\begin{align}
  \pi\sum_{f=1}^2\lambda_f^2&\int^R dr\; r\left(|\phi_f|^2 - v_f^2\right)^2 \sim
  \textrm{finite}, \\
  \int^R dr\; r\int_0^{2\pi}d\theta&\left[\eta-\gamma\left(\phi_1^{Q_2}\bar{\phi}_2^{Q_1} + \bar{\phi}_1^{Q_2}\phi_2^{Q_1}\right)\right]\non
  &\sim \int^R dr\; r\int_0^{2\pi}d\theta\left[\eta-2\gamma
    v_1^{Q_2}v_2^{Q_1}\cos(n_{\rm global}\Theta_{\rm global})\right]\non
  &\sim 4\gamma v_1^{Q_2}v_2^{Q_1}
  \int^R dr\; r\int_0^{\frac{2\pi}{n_{\rm global}}}d\theta\;r
  \sum_{\mathsf{m}=0}^{n_{\rm global}-1}
  \sech^2\left[(\theta - \tilde{\vartheta}_{\mathsf{m}})r\kappa\right]\non
  &\sim
  \frac{8\sqrt{\gamma}v_1^{\frac{Q_2}{2}+1}v_2^{\frac{Q_1}{2}+1}n_{\rm global}}
  {\sqrt{Q_1^2v_1^2 + Q_2^2v_2^2}} R.
\end{align}
Summing up all contributions, we have a linearly divergent energy
\beq
E =
\frac{16\sqrt{\gamma}v_1^{\frac{Q_2}{2}+1}v_2^{\frac{Q_1}{2}+1}n_{\rm global}}
     {\sqrt{Q_1^2v_1^2 + Q_2^2v_2^2}} R
     + \mathcal{O}\big(\log R, R^0\big),
\eeq
if $\gamma\neq 0$ \emph{and} $n_{\rm global}\neq 0$.

Although it is clear that for $\gamma=0$ and $n_{\rm global}\neq 0$,
there is a logarithmic divergence, our analysis does not show whether
there is a subleading logarithmic divergence in the case of
$\gamma\neq 0$ and $n_{\rm global}\neq 0$; 
that is, however, beyond the scope of this paper.

\section{Numerical results}\label{sec:results}

\subsection{Numerical method}\label{sec:num_method}

Our numerical calculations are carried out using a custom built CUDA C
code for an NVIDIA GPU cluster and the code uses a simple gradient
flow method to find the vortex solutions in the two-flavor
Abelian-Higgs systems with a generalized Josephson interaction term. 
The static equations of motion are discretized using a fourth-order
5-point stencil of a standard finite difference scheme and the lattice
points are updated using the fourth-order Runge-Kutta method.

\subsection{Visualization}\label{sec:vis}

\begin{figure}[!htp]
  \begin{center}
    \mbox{\subfloat[]{\includegraphics[width=0.45\linewidth]{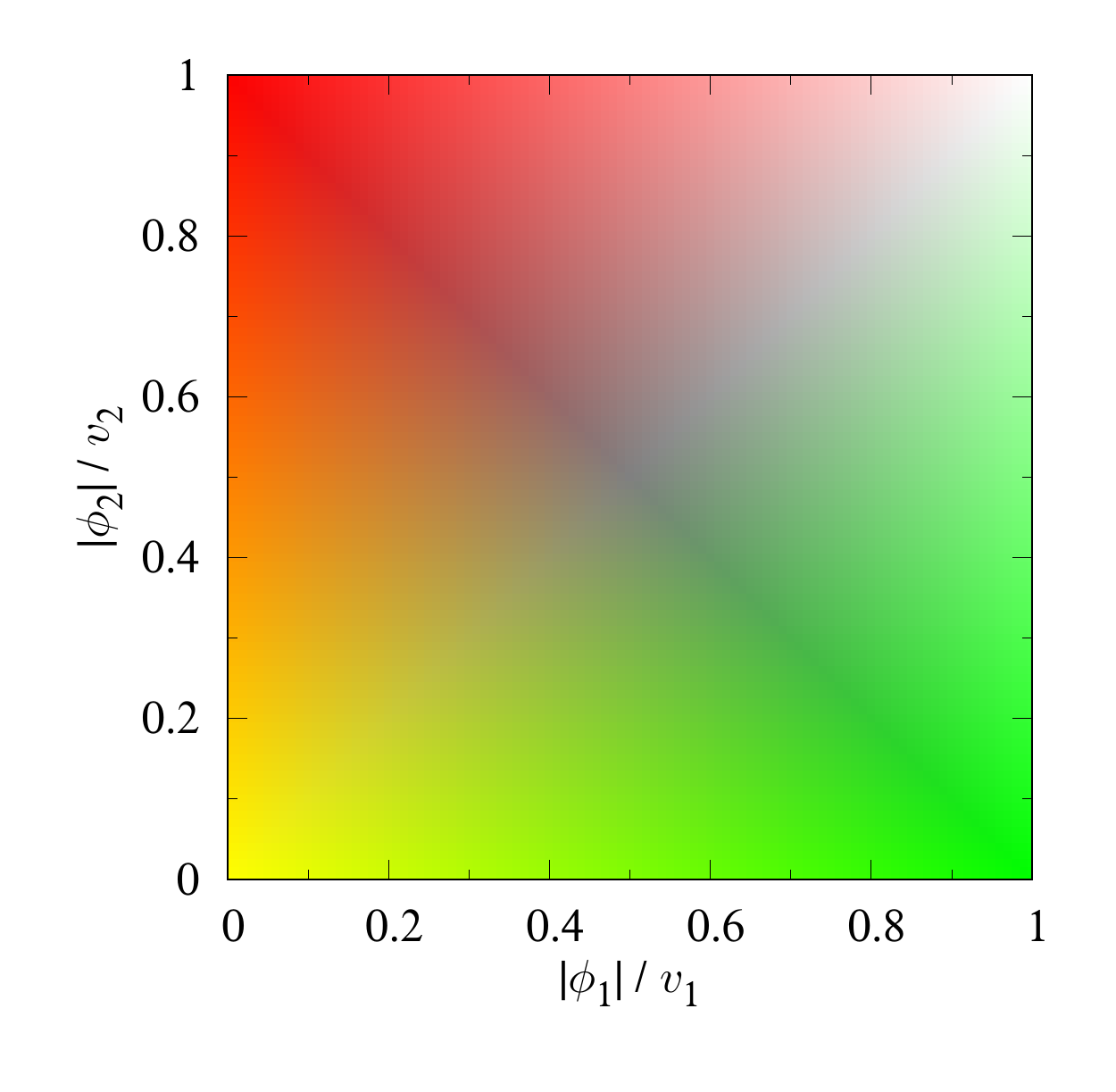}}
      \subfloat[]{\includegraphics[width=0.45\linewidth]{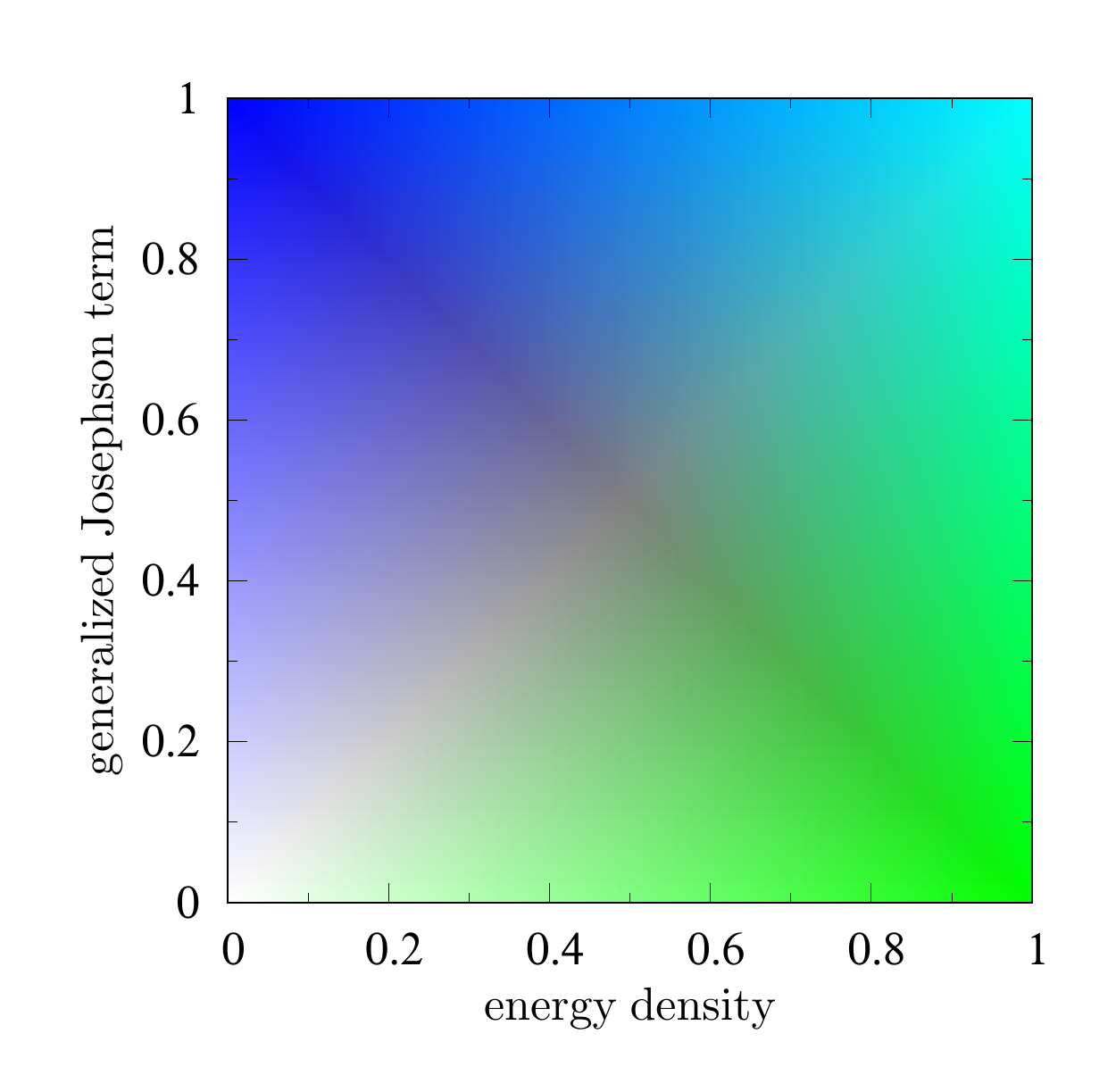}}}
    \caption{(a) Visualization of 2 complex fields using a customized
      coloring scheme. The vacuum is displayed with white, the
      vortices of $\phi_1$ with red, the vortices of $\phi_2$ with
      green and the coincident vortices of both $\phi_1$ and $\phi_2$
      with yellow. (b) Visualization of the energy density and the
      generalized Josephson term (density) using a customized coloring
      scheme. The vacuum is displayed with white, the pure energy with
      green, the pure generalized Josephson term with blue and the
      overlap of the latter two with cyan. }
    \label{fig:color_scheme}
  \end{center}
\end{figure}

In order to visualize the results, we will employ a coloring scheme
illustrated in fig.~\ref{fig:color_scheme}.
The absolute values of the two complex scalar fields, $|\phi_{1,2}|$,
are used via a 2-dimensional coloring scheme illustrated in the figure
to visualize the field configurations.
The scheme is constructed such that the vacuum ($|\phi_1|=v_1$ and
$|\phi_2|=v_2$) is white, whereas the position of the vortices in
$\phi_1$ (i.e.~$\phi_1\approx 0$) are displayed with red and vortices
in $\phi_2$ (i.e.~$\phi_2\approx 0$) are displayed with green.
The scheme is made such that the superposed vortices become yellow and
the intermediate values of both fields turns out to become an
interpolation function between red and green that goes through gray,
see fig.~\ref{fig:color_scheme}. 

\subsection{An intuitive explanation}\label{sec:intuitive}

We will first and foremost be interested in local and hence
finite-energy vortices, which poses the constraint $Q_2n_1=Q_1n_2$
(i.e.~$n_{\rm global}=0$) and in turn the winding number compensated
by the gauge field is not just fractional, but rational, see
eq.~\eqref{eq:rational_k}. 

\begin{figure}[!htp]
  \begin{center}
    \mbox{\subfloat[]{\includegraphics[width=0.2\linewidth]{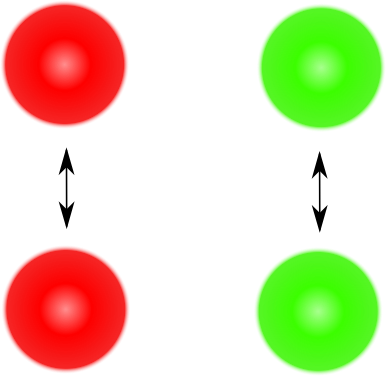}}\qquad
      \subfloat[]{\includegraphics[width=0.4\linewidth]{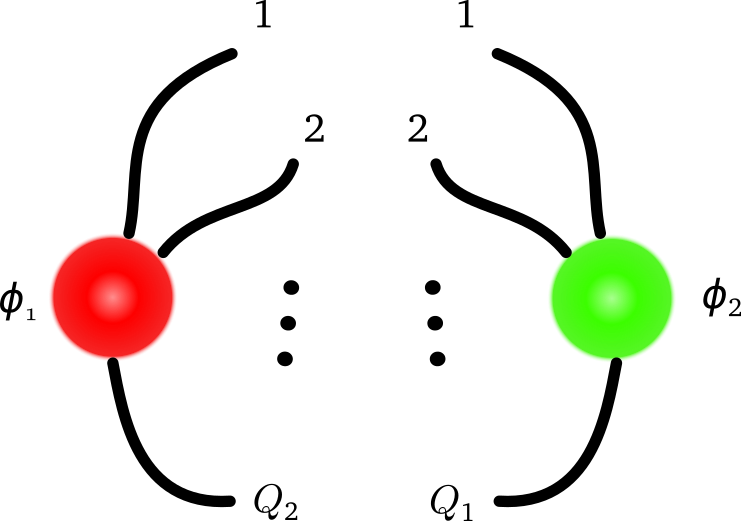}}\qquad
      \subfloat[]{\includegraphics[width=0.25\linewidth]{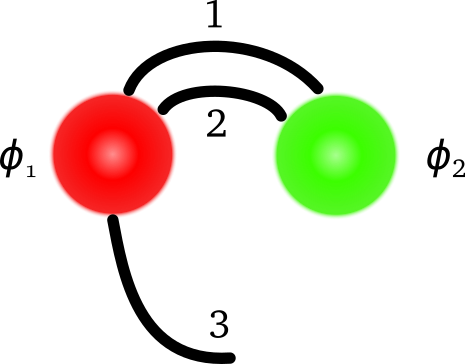}}}
    \caption{Intuitive explanation of building a vortex solution in
      our model.
      (a) Vortices of the same species repel each other,
      (b) but for $\gamma\neq 0$, $\phi_1$ has $Q_2$ legs while
      $\phi_2$ has $Q_1$ legs.
      (c) The connected legs attract the two vortices of opposite
      species in order to minimize the angular domain wall length.
      If there is a leg that cannot be connected anywhere, then
      $n_{\rm global}\neq 0$: in this example $Q_2=3$, $Q_1=2$ and
      $n_1=n_2=1$ so $n_{\rm global}=Q_2-Q_1=1\neq 0$.
    }
    \label{fig:toymodel}
  \end{center}
\end{figure}

First we should mention that we will consider only the strong type II
regime for both vortex species, which is tantamount to the conditions
\beq
\lambda_f \gg e, \qquad f=1,2.
\eeq
This means that vortices of the \emph{same} species will repel each
other, but vortices with different species only interact with each
other via the common gauge field and the generalized Josephson term
($\gamma\neq 0$), see fig.~\ref{fig:toymodel}(a). 

In order to understand what the solutions of the model with
$\gamma\neq 0$ look like, it will be illuminating to introduce a
simple toy model, where each vortex in the field $\phi_1$ has $Q_2$
legs -- corresponding to emanating angular domain walls -- and each
vortex of the field $\phi_2$ has $Q_1$ legs, see
fig.~\ref{fig:toymodel}(b).

In fig.~\ref{fig:toymodel}(c), we have made a simple example where the
charges are assigned as $Q_2=3$ and $Q_1=2$, so that the red vortex
species ($f=1$) has three legs and the green vortex species ($f=2$)
has two legs.
We can only connect legs from one vortex species to the opposite
species.
If any legs remain, the global vortex number is nonvanishing: the
total energy will thus be linearly diverging, as demonstrated in
sec.~\ref{sec:adw}. 
In this simple example, there is one leg from the red vortex species
remaining; thus the global vortex number, $n_{\rm global}=1$.
Had the remaining leg been emanating from the green vortex species, we
should count it with a negative sign in the global vortex number and
hence $n_{\rm global}$ would be minus one instead.

\begin{enumerate}
\item[]\emph{Summary of how to construct vortices in our model:}
\item Place $n_1$ red vortices and $n_2$ green vortices, each with
  $Q_2$ and $Q_1$ legs, respectively.
\item Connect as many legs between two opposite vortex species as
  possible. 
  \begin{enumerate}
  \item If no legs remain, we have successfully constructed a
    \emph{local} vortex.
  \item If legs are remaining, the global vortex number is given by
    the number of legs from the red vortex species ($f=1$) or
    \emph{minus} the number of legs from the green vortex species
    ($f=2$). 
  \end{enumerate}
\end{enumerate}

\subsection{Constructing local vortices}\label{sec:constructing_local}

We will almost exclusively consider local vortex solutions in this
paper, which entails solutions that obey
\beq
n_{\rm global}
= Q_2 n_1 - Q_1 n_2 = 0,
\eeq
for which the ``local'' vortex number must be
\beq
n_{\rm local} = 2Q_2 n_1 = 2Q_1 n_2,
\eeq
and in turn the winding number compensated by the gauge field is given
by eq.~\eqref{eq:rational_k}.
The local vortex number is just a mathematically convenient construct,
whereas physically the winding number $k$ is much more intuitive.

The minimally winding local vortices -- which have finite energy and
are spatially localized, have vortex numbers
\beq
n_{1,{\rm min}} = \frac{Q_1}{\gcd(Q_1,Q_2)}, \qquad
n_{2,{\rm min}} = \frac{Q_2}{\gcd(Q_1,Q_2)},
\eeq
for which the local vortex number is
\beq
n_{\rm local, min} = \frac{2Q_1Q_2}{\gcd(Q_1,Q_2)},
\eeq
while the winding number is
\beq
k_{\rm min} = \frac{1}{\gcd(Q_1,Q_2)}.
\eeq
Obviously, we can have $n$ copies of such minimal local vortices,
which just amounts to multiplying the above quantities by the integer
$n\in\mathbb{Z}_{>0}$.

\subsection{Minimal solutions}\label{sec:minimal_local}

We will be interested only in the strong type-II domain where
\beq
\frac{m_{\phi_f}}{m_\gamma} = \frac{\sqrt{2}\lambda_f v_f}{e\sqrt{Q_1^2v_1^2 + Q_2^2v_2^2}} \gg 1, \qquad\textrm{($f$ not summed over)}
\eeq
for both $f=1,2$.
For definiteness, we will choose
\begin{align}
  e&=0.3,\\
  \lambda_1=\lambda_2&=2.5,\\
  \gamma&=0.1,
\end{align}
throughout the paper for all numerical calculations and also
\beq
v_1=v_2=1,
\eeq
in this section, but we change their ratio in
sec.~\ref{sec:varying_vevs}. 
In order to retain stability of the vortex vacuum, we only consider
the generalized Josephson term as a perturbation and hence take
$\gamma\ll\lambda_{1,2}$, for which $\gamma=0.1$ is reasonable.

In this section, we will study the minimal solutions, by which we mean
the minimal vortex numbers possible for a \emph{local} vortex
solution.
This will in turn also result in the lowest energy solution --
although for charges larger than 3, we begin to find energetically
metastable solutions with the same vortex numbers as the global
energy minimizing solution.

\subsubsection{\texorpdfstring{$Q_1=Q_2=Q$}{Q1=Q2=Q} : The normal vortex}

The simplest case is when the two charges are equal.
This means that the greatest common divisor is $\gcd(Q_1,Q_2)=Q$ and
for $Q>1$ we have rational (fractional) vortices with
$k=\frac{1}{Q}$. 
Nevertheless, since the two vortex species have the same number of
legs, the minimal local vortex is simply given by connecting the $Q$
legs between two vortices of each species. 

\def\figsize{0.3}
\begin{figure}[!htp]
  \begin{center}
    \mbox{\subfloat[$(Q_1,Q_2)=(1,1)$]{\includegraphics[scale=\figsize]{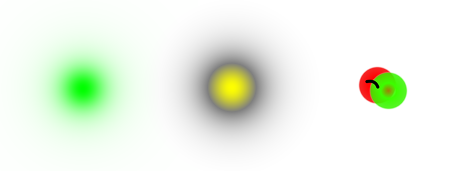}}\qquad
      \subfloat[$(Q_1,Q_2)=(2,2)$]{\includegraphics[scale=\figsize]{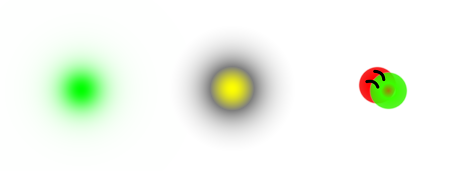}}}
    \mbox{\subfloat[$(Q_1,Q_2)=(3,3)$]{\includegraphics[scale=\figsize]{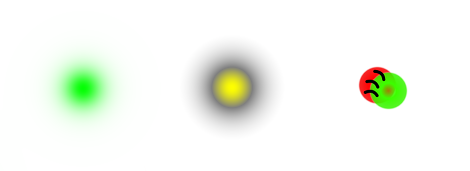}}\qquad
      \subfloat[$(Q_1,Q_2)=(4,4)$]{\includegraphics[scale=\figsize]{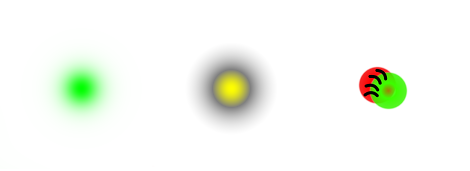}}}
    \caption{$Q_1=Q_2=Q$ minimal ``normal'' vortices with
      $Q=1,2,3,4$. Each subfigure contains 3 elements from left to
      right: an energy density plot, a field density plot and a sketch
    of how the red and green vortices are interconnected. The first
    two plots are made using the coloring scheme of
    fig.~\ref{fig:color_scheme}.} 
    \label{fig:minimal_normal}
  \end{center}
\end{figure}

Fig.~\ref{fig:minimal_normal} shows the minimal ``normal'' vortices
for $Q=1,2,3,4$ and they are all just a pair of a vortex of each of
the two species.
Each subfigure shows the energy density and the field densities with
the color schemes of fig.~\ref{fig:color_scheme} as well as a sketch
of the vortex components and how they are connected with their legs.

The solutions thus look very similar and in fact the only difference
between the solutions is the effective photon mass
\beq
m_\gamma = e Q \sqrt{v_1^2 + v_2^2},
\eeq
which changes with $Q$ and this in turn can be seen in
fig.~\ref{fig:minimal_normal} as different interpolations between
yellow and gray (and then white being the vacuum). 

The nontriviality of the local solutions will be seen first when the
charges differ from unity and from each other; we will classify them
in various categories next.

\subsubsection{\texorpdfstring{$(Q_1,Q_2)=(Q,1)$}{(Q1,Q2)=(Q,1)} : The
  vortex flower}

The simplest nontrivial minimal vortex exists in the theory with one
of the charges equal to unity and the other larger: Thus we can take
$(Q_1,Q_2)=(Q,1)$ for $Q>1$.
The vortex of the second species (green) will now have $Q$ legs and
thus naturally be positioned in the center of the local vortex with
$Q$ vortices of the first species (red) as petals around a receptacle. 

\begin{figure}[!htp]
  \begin{center}
    \mbox{\subfloat[$(Q_1,Q_2)=(2,1)$]{\includegraphics[scale=\figsize]{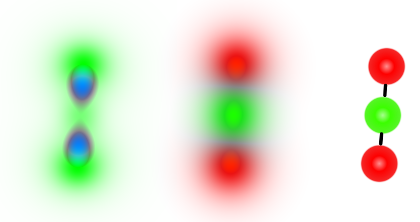}}\qquad\qquad
      \subfloat[$(Q_1,Q_2)=(3,1)$]{\includegraphics[scale=\figsize]{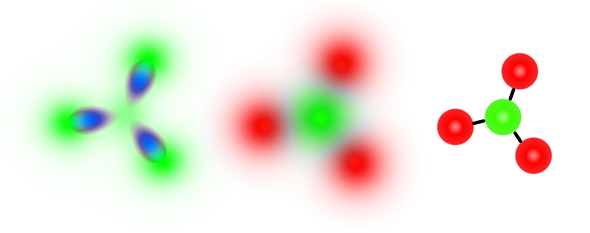}}}
    \mbox{\subfloat[$(Q_1,Q_2)=(4,1)$]{\includegraphics[scale=\figsize]{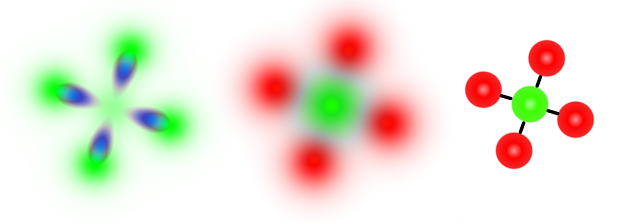}}}
    \caption{$(Q_1,Q_2)=(Q,1)$ minimal vortex ``flowers'', with
      $Q=2,3,4$. Each subfigure contains 3 elements from left to
      right: an energy density plot, a field density plot and a sketch
    of how the red and green vortices are interconnected. The first
    two plots are made using the coloring scheme of
    fig.~\ref{fig:color_scheme}.}
    \label{fig:minimal_flowers}
  \end{center}
\end{figure}

Fig.~\ref{fig:minimal_flowers} displays the minimal ``flower''
vortices for $Q=2,3,4$.
Each subfigure shows the energy density with the density of the
generalized Josephson term overlaid using the color scheme of
fig.~\ref{fig:color_scheme}(b), the field densities using the color
scheme of fig.~\ref{fig:color_scheme}(a) as well as a sketch of the
vortex components and how their legs are connecting them. 
As anticipated, the solutions look like flowers with $Q$ petals on a
single receptacle.
The legs are clearly visible on energy part of the subfigures as
overlaid density of the generalized Josephson term.

One may speculate what if we were to increase $Q$: Would the solution
still look like a flower? First of all, if $Q$ is larger than 4,
stability is lost for any value of $\gamma$, as a runaway direction is
opened up. The second issue is the bond length of the petal to the
receptacle being model dependent; meaning that it depends on the values
of $\lambda_f$ and $e$, etc.
Some exploration revealed that what happens if there is energetically
no room for an extra petal, the solution turns into a flower with
$Q-1$ petals and a vortex with both species as the receptacle.

\subsubsection{\texorpdfstring{$(Q_1,Q_2)=(Q+1,Q)$}{(Q1,Q2)=(Q+1,Q)} : The
vortex stick}

The next class of minimal vortices appear for $(Q_1,Q_2)=(Q+1,Q)$ with
$Q\geq 1$: We call this class of solutions vortex ``sticks''.
The solution looks like a stick with $Q+1$ red vortices and $Q$ green
vortices of alternating colors.
Notice that for $Q=1$, the solution is exactly the same as the vortex
``flower'' with two petals. 

\begin{figure}[!htp]
  \begin{center}
    \mbox{\subfloat[$(Q_1,Q_2)=(2,1)$]{\includegraphics[scale=\figsize]{vortex_2_1_out}}\qquad\qquad
      \subfloat[$(Q_1,Q_2)=(3,2)$]{\includegraphics[scale=\figsize]{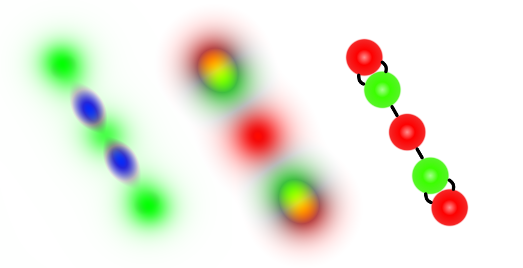}}}
    \mbox{\subfloat[$(Q_1,Q_2)=(4,3)$]{\includegraphics[scale=\figsize]{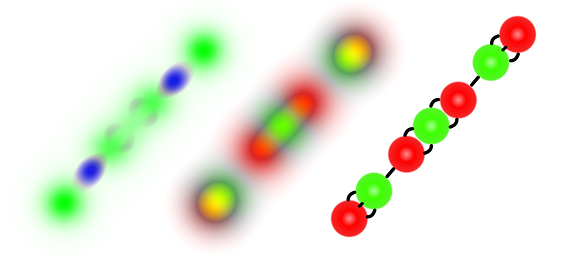}}}
    \caption{$(Q_1,Q_2)=(Q+1,Q)$ vortex ``sticks'', with
      $Q=1,2,3$. Each subfigure contains 3 elements from left to 
      right: an energy density plot, a field density plot and a sketch
    of how the red and green vortices are interconnected. The first
    two plots are made using the coloring scheme of
    fig.~\ref{fig:color_scheme}.}
    \label{fig:minimal_sticks}
  \end{center}
\end{figure}

Fig.~\ref{fig:minimal_sticks} shows the vortex ``sticks'' with
$Q=1,2,3$.
Each subfigure shows the energy densities with overlaid density of the
generalized Josephson term, the field densities as well as a sketch of
the leg connections. 
For $Q=3$, this is not the only possibility as we will see shortly.

For this class, we see for the first time the situation where more
than one leg is connecting two adjacent vortices of opposite
species.
In fig.~\ref{fig:minimal_sticks}(b), the double ``bond'' appears
twice and is artistically drawn on the sketch on the sides of the two
vortices; nevertheless, the overlay of the density of the
generalized Josephson term clearly shows that the binding ``bond''
does indeed look like two separated legs.
We should also notice that the length of the ``bonds'' for the double
bond case is shorter than in the single-bond case.
This trend continues to the triple-bond case where, unfortunately, it
becomes impossible to see the density of the generalized Josephson
term overlaying the total energy density.
This is because the bond length is extremely short and the energy
density is locally too small compared to that of the single bond.

\subsubsection{\texorpdfstring{$(Q_1,Q_2)=(2Q,Q)$}{(Q1,Q2)=(2Q,Q)} :
  The vortex pill}

As we cannot take the charges to be too large, we have not yet seen an
example where the charges have a common factor.
The simplest case and the only possible one for charges less-or-equal
to four is the case of $(Q_1,Q_2)=(4,2)$, which thus can attain the
solution of the $(Q_1,Q_2)=(2,1)$ model by doubling the number of legs
between the vortices.
We will call this type, the vortex ``pill'', which is nothing but the
simplest vortex ``flower'' with only two petals or the simplest vortex
``stick''. 

\begin{figure}[!htp]
  \begin{center}
    \mbox{\subfloat[$(Q_1,Q_2)=(2,1)$]{\includegraphics[scale=\figsize]{vortex_2_1_out}}\qquad\qquad
      \subfloat[$(Q_1,Q_2)=(4,2)$]{\includegraphics[scale=\figsize]{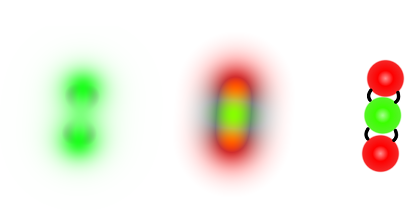}}}
    \caption{$(Q_1,Q_2)=(2Q,Q)$ minimal vortex ``pills'', with
      $Q=1,2$. Each subfigure contains 3 elements from left to 
      right: an energy density plot, a field density plot and a sketch
    of how the red and green vortices are interconnected. The first
    two plots are made using the coloring scheme of
    fig.~\ref{fig:color_scheme}.}
    \label{fig:minimal_pills}
  \end{center}
\end{figure}

Fig.~\ref{fig:minimal_pills} shows the minimal vortex ``pills'' for
$Q=1,2$.
Each subfigure shows the energy density with overlaid density of the
generalized Josephson term, the field densities as well as a sketch of
the leg connections.
The double bonds in fig.~\ref{fig:minimal_pills}(b) versus the single
bonds in fig.~\ref{fig:minimal_pills}(a) are clearly visible in the
energy part of the subfigures.

\subsubsection{\texorpdfstring{$(Q_1,Q_2)=(4,3)$}{(Q1,Q2)=(4,3)} : The
  extended flower}

So far all the minimal vortices have been the \emph{local} vortex
solutions with the smallest possible vortex numbers for given
electric charges.
Additionally, they have also been the global energy minimizing
solutions for all cases but one -- namely the $(Q_1,Q_2)=(4,3)$ case:
In this case, the vortex ``stick'' is not the energetically favorable
solution, but a slightly lower total energy can be obtained by
deforming the stick into a ``flower'' form. 

\begin{figure}[!htp]
  \begin{center}
    \mbox{\includegraphics[scale=\figsize]{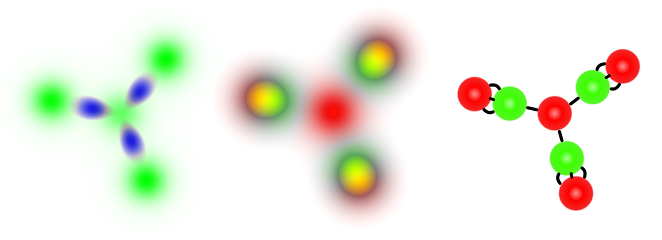}}
    \caption{$(Q_1,Q_2)=(4,3)$ minimal vortex ``extended flower'', which is
      the minimum-energy local vortex in its charge sector. 
      The figure contains 3 elements from left to 
      right: an energy density plot, a field density plot and a sketch
    of how the red and green vortices are interconnected. The first
    two plots are made using the coloring scheme of
    fig.~\ref{fig:color_scheme}.}
    \label{fig:minimal_extended_flower}
  \end{center}
\end{figure}

Fig.~\ref{fig:minimal_extended_flower} shows the lowest-energy
$(Q_1,Q_2)=(4,3)$ local vortex, which takes the form of a flower with
composite petals on a single-vortex receptacle.
Each subfigure shows the energy density with overlaid density of the
generalized Josephson term, the field densities as well as a sketch of
the leg connections.
The composite petals are made of two vortices -- one of each species
-- and they are connected internally by a triple bond.
Unfortunately, the short triple bond is not visible on the energy part
of the figure, whereas the single bonds are very clearly marked in the
overlaid density of the generalized Josephson term.
Although the composite petal is very compact, it is nevertheless
possible to see on the field densities part of the figure that the
green vortex is the inner-most one as it must connect to the red
central receptacle.

\subsection{Nonminimal solutions}\label{sec:nonminimal_local}

It should be obvious by now that the minimal vortices cannot be the
only possibilities of connecting the legs to form local vortices.
We call the remaining infinity of solutions, nonminimal solutions.
Indeed as there are infinitely many nonminimal solutions, we will only
display a selected few solutions that we have found numerically.

\begin{figure}[!htp]
  \begin{center}
    \mbox{\includegraphics[scale=\figsize]{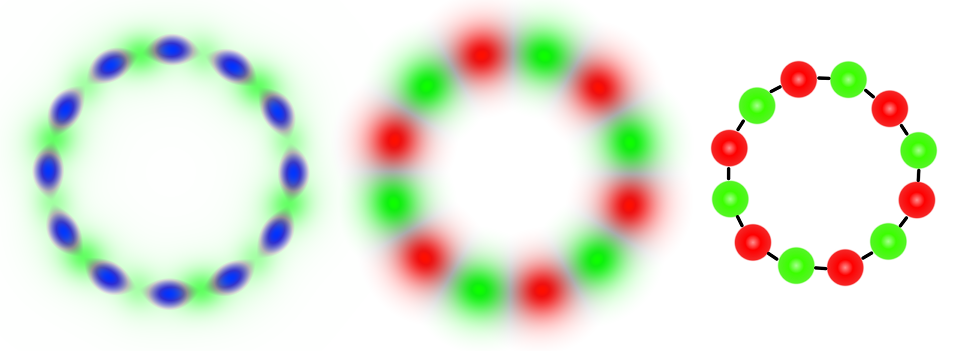}}
    \caption{$(Q_1,Q_2)=(2,2)$ nonminimal vortex ring. The figure
      contains 3 elements from left to 
      right: an energy density plot, a field density plot and a sketch
    of how the red and green vortices are interconnected. The first
    two plots are made using the coloring scheme of
    fig.~\ref{fig:color_scheme}.}
    \label{fig:nonminimal22}
  \end{center}
\end{figure}

Fig.~\ref{fig:nonminimal22} shows the case with the smallest possible
electric charges possessing nonminimal solutions, viz.~the
$(Q_1,Q_2)=(2,2)$ case.
The minimal local vortex with these charges is simply a vortex of both
species superposed one on the other, see
fig.~\ref{fig:minimal_normal}(b).
Instead of using two legs to connect the two vortices of opposite
species, we can connect two green vortices to a single red vortex,
which leaves one leg unconnected.
The solution is simple:
Indeed we can make a ring network of vortices of alternating species
(red, green, red, $\ldots$, green).
The particular ring network solution found and illustrated in
fig.~\ref{fig:nonminimal22} contains six vortices of each species.

Although the red vortex and the green vortex (i.e.~of species 1 and
species 2, respectively) do not repel each other, the strong repulsive
force between vortices of the same species push them as far apart as
possible, turning the solution into a network of alternating vortex
species on a wound-up string -- a ring network.
The reason for the whole construction not to fall apart, is that the
legs contribute to the energy via the generalized Josephson term and
hence must be shortened as much as possible.
The balance act between the two opposing forces creates the bond
length and in turn the specific solution depicted in the figure.

\begin{figure}[!htp]
  \begin{center}
    \mbox{\includegraphics[scale=\figsize]{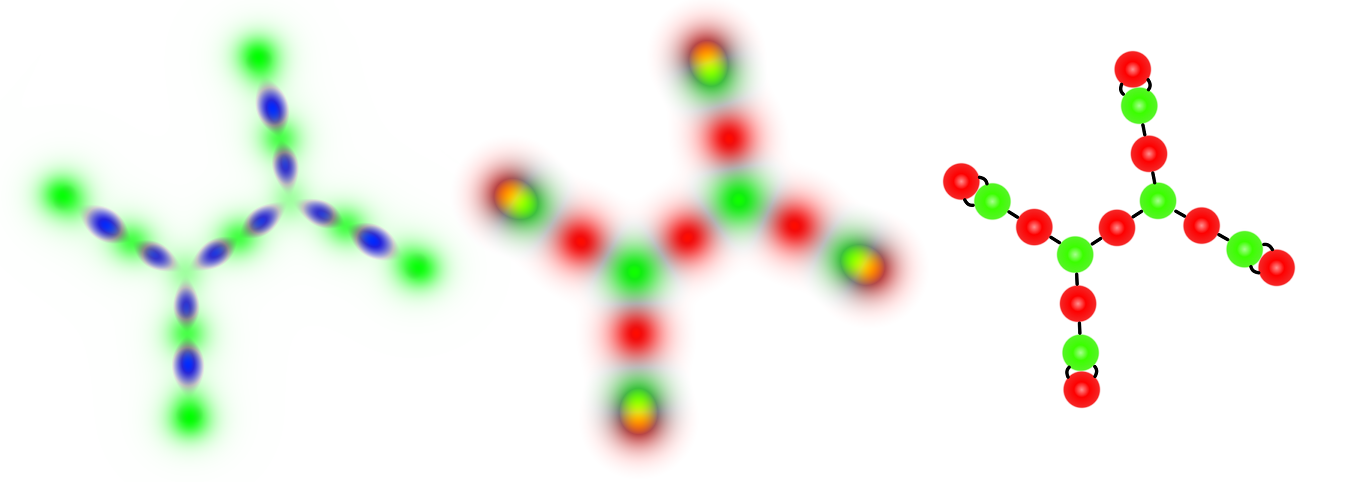}}
    \caption{$(Q_1,Q_2)=(3,2)$ nonminimal vortex stick network. 
      The figure contains 3 elements from left to 
      right: an energy density plot, a field density plot and a sketch
    of how the red and green vortices are interconnected. The first
    two plots are made using the coloring scheme of
    fig.~\ref{fig:color_scheme}.}
    \label{fig:nonminimal32}
  \end{center}
\end{figure}

The next example of a nonminimal local vortex is for the
$(Q_1,Q_2)=(3,2)$ case.
The minimal local vortex with these charges is the vortex ``stick'',
see fig.~\ref{fig:minimal_sticks}(b).
The $Q=2$ stick (i.e.~$(Q_1,Q_2)=(Q+1,Q)=(3,2)$) has two parts with
double bonds as opposed to the $Q=1$ stick, which has only single
bonds.
The double bonds can thus be split and used to connect other sticks
and it is then unlimited how large a stick network one can build from
this foundation.

The two basic building blocks in this construction are a vertex
composed by a green vortex as a receptacle with three red petals, each
having an unconnected leg, and the other is an end cap composed by a
red and a green vortex interconnected by a double bond, leaving the
green vortex with an unconnected leg. 
The two types of bond are clearly visible in
fig.~\ref{fig:nonminimal32} in the energy density part of the figure
(the left-most part).

The specific vortex stick network depicted in
fig.~\ref{fig:nonminimal32} is composed by three minimal vortex sticks 
that have been deformed so as to form the stick network resembling the
capital Latin letter `H'. 

\begin{figure}[!htp]
  \begin{center}
    \mbox{\includegraphics[scale=\figsize]{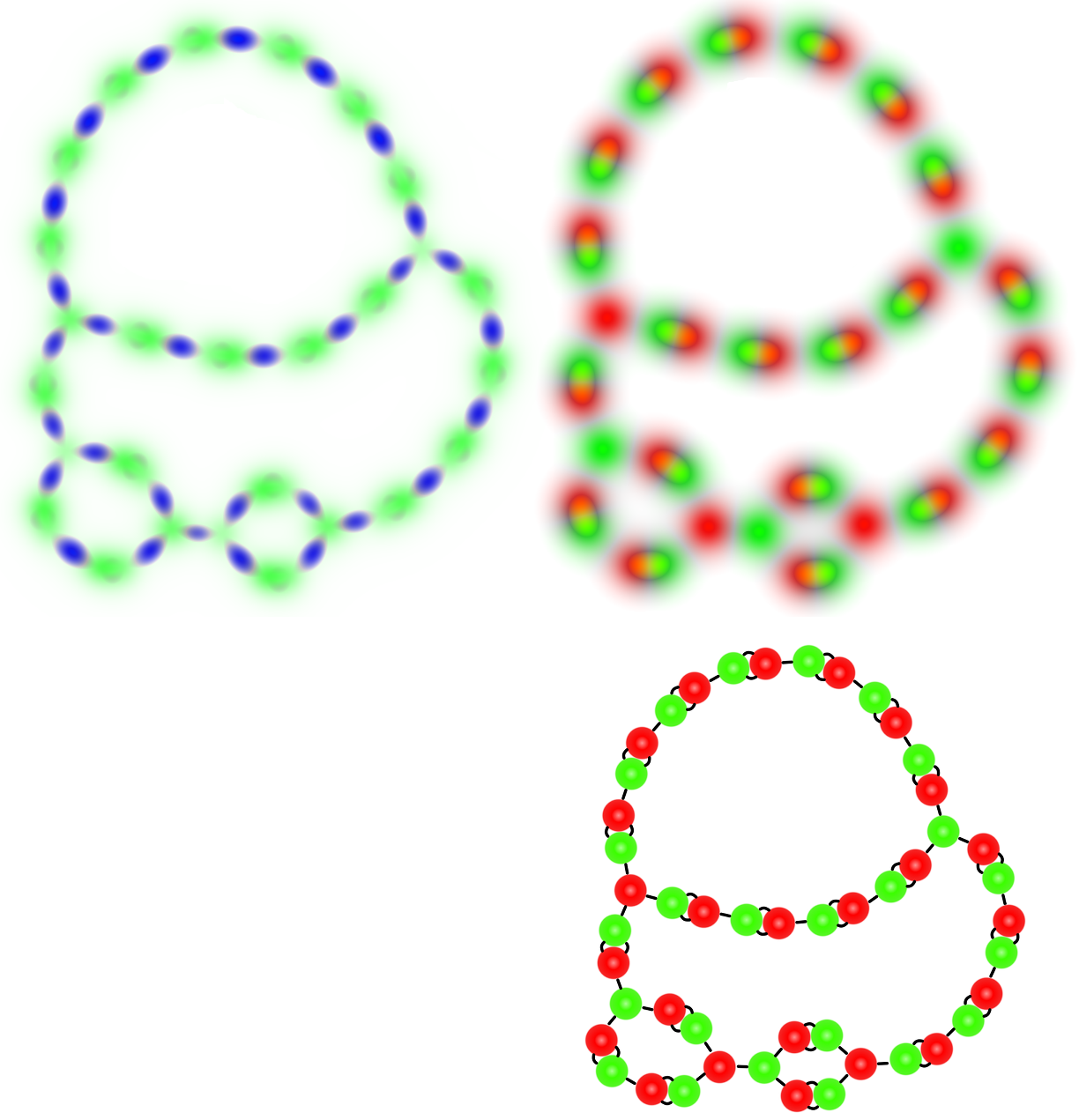}}
    \caption{$(Q_1,Q_2)=(3,3)$ nonminimal vortex ring network. 
    The figure contains 3 elements from left to right and then down:
    an energy density plot, a field density plot and a sketch 
    of how the red and green vortices are interconnected. The first
    two plots are made using the coloring scheme of
    fig.~\ref{fig:color_scheme}.}
    \label{fig:nonminimal33}
  \end{center}
\end{figure}

The next example of a nonminimal local vortex is for the
$(Q_1,Q_2)=(3,3)$ case.
Since the two electric charges are equal, the minimal vortex is simply
a vortex of each species, mutually interconnected by three legs, see
fig.~\ref{fig:minimal_normal}(c). 
This triple bond can be split into a double bond with two loose legs
or into a single bond with four loose legs.
This complexity yields vortex ring network solutions with two basic
components.
The mentioned former option provides a link composed by a red and a
green with a free leg from each one of them.
The other basic component is the vertex, which is either a single red
or a single green vortex with three legs connecting three chains of
links.

Fig.~\ref{fig:nonminimal33} shows the vortex ring network solution
that we found numerically.
Notice that at every vertex, the incoming legs must (of course) all be
from the opposite vortex species, which in turn makes the chains of
links directed graphs.
It is definitely possible to make a simple ring out of a linked chain,
which would be a simpler solution than the one depicted in
fig.~\ref{fig:nonminimal33}.
The above observation implies that once a vertex is inserted into a
ring, an even number of vertices must be inserted in order to close
the ring and hence make the vortex local. 
The double bonds in the links and the three single bond emanating from
the vertices are clearly visible in blue on the energy density part of
fig.~\ref{fig:nonminimal33}. 

\begin{figure}[!htp]
  \begin{center}
    \mbox{\includegraphics[scale=\figsize]{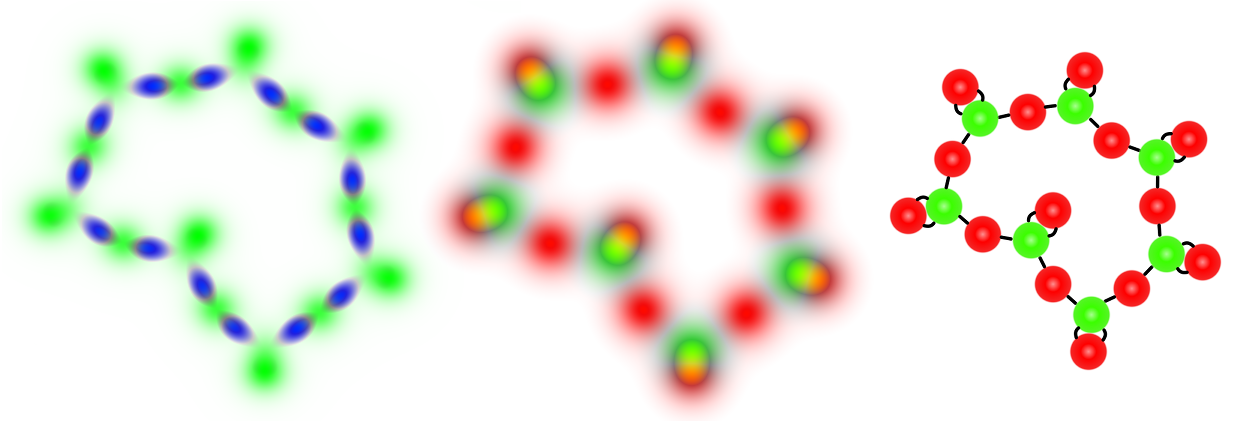}}
    \caption{$(Q_1,Q_2)=(4,2)$ nonminimal vortex ring network.
    The figure contains 3 elements from left to 
      right: an energy density plot, a field density plot and a sketch
    of how the red and green vortices are interconnected. The first
    two plots are made using the coloring scheme of
    fig.~\ref{fig:color_scheme}.}
    \label{fig:nonminimal42}
  \end{center}
\end{figure}

The next example is the $(Q_1,Q_2)=(4,2)$ case, whose minimal local
vortex solution is in the class of vortex ``pills'', see
fig.~\ref{fig:minimal_pills}(b).
The minimal local vortex has two double bonds and either double bond
can be split into two single bonds making the green vortex a link
piece with a red dweller on its back.
There are thus two basic components in this model, the green link
piece with a red dweller and simply a red vortex.
Alternating these two components makes it possible to create a vortex
ring network of any size.

Fig.~\ref{fig:nonminimal42} shows a nonminimal vortex ring with seven
links of both types described above.
Since the dwelling red vortex is attached to the green, necessarily on
one side of the green vortex (due to repulsion from the other red
vortices in the ring network), the red dweller can either sit on the
inside or the outside of the ring.
The link naturally curves the ring away from the dweller and hence
most dwellers are situated on the outside, since the overall curvature
of the circle must be positive for the ring to close.
Both the double bonds associated with the dweller and the single bond
connecting the two types of link in the ring network of vortices are
visible in the energy density part of fig.~\ref{fig:nonminimal42}. 

\begin{figure}[!htp]
  \begin{center}
    \mbox{\includegraphics[scale=\figsize]{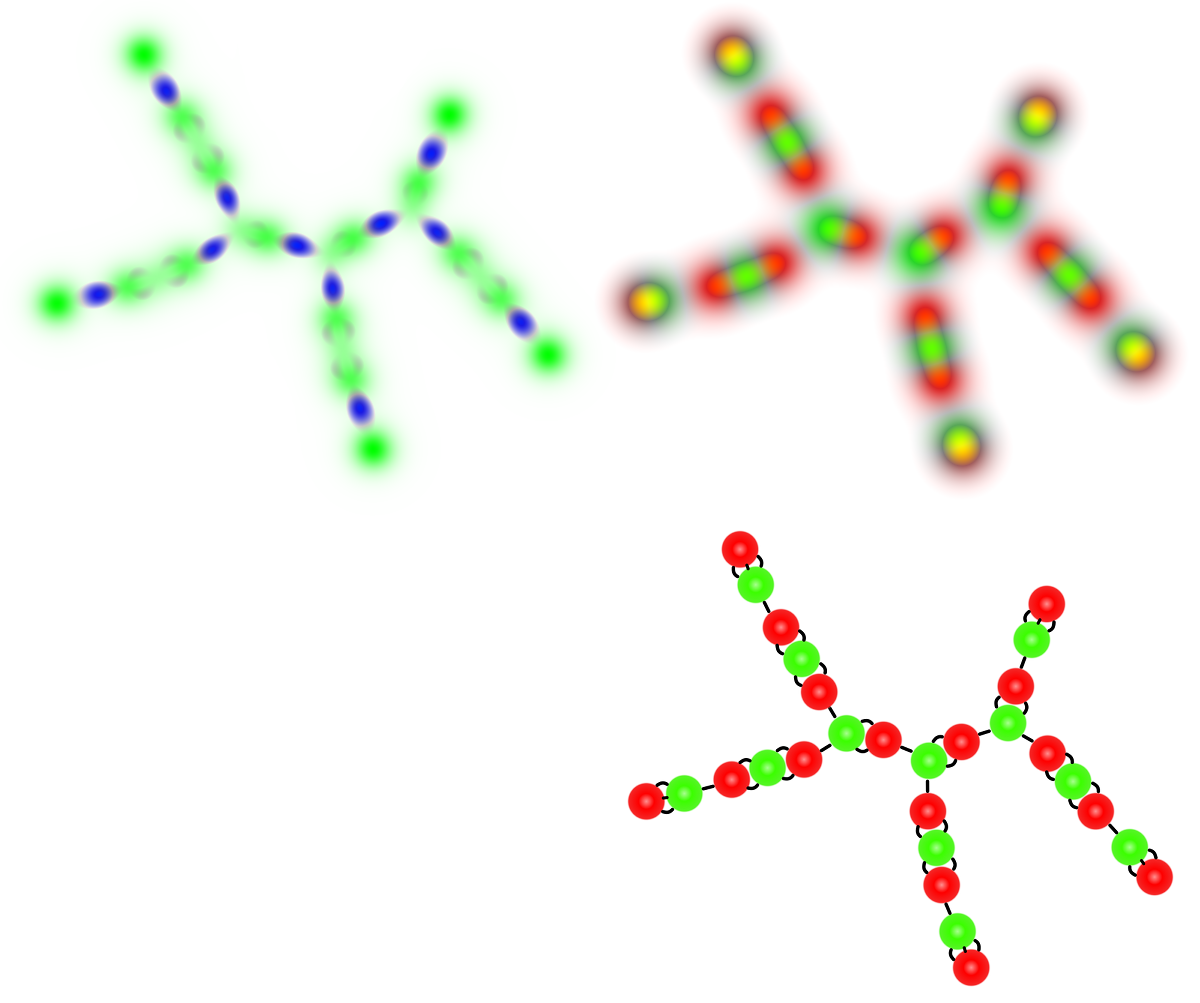}}
    \caption{$(Q_1,Q_2)=(4,3)$ nonminimal vortex stick network. 
    The figure contains 3 elements from left to right and then down:
    an energy density plot, a field density plot and a sketch 
    of how the red and green vortices are interconnected. The first
    two plots are made using the coloring scheme of
    fig.~\ref{fig:color_scheme}.}
    \label{fig:nonminimal43}
  \end{center}
\end{figure}

The next example is the $(Q_1,Q_2)=(4,3)$ case, whose minimal local
vortex solutions are a ``stick'', see fig.~\ref{fig:minimal_sticks}(c)
and the ``extended flower'', see
fig.~\ref{fig:minimal_extended_flower} -- the latter of which has the
lowest total energy. 
The $Q=3$ stick (i.e.~$(Q_1,Q_2)=(4,3)$) has two triple bonds, two
double bonds and two single bonds.
Both the triple and double bonds can be split making the stick
attachable to other sticks and thus provides the breeding ground for
the vortex stick network.

Three basic building blocks can be made: An end cap which is a green
and a red vortex interconnected by a triple bond, leaving the green
vortex with a loose leg.
The second building block is a link, which is made of two red vortices
encapsulating a single green vortex with two double bonds, leaving
both red vortices with a free leg each.
The third and final building block is the vertex, which comes in two
flavors: a three- and four-legged vertex; the formed is made by
attaching a red vortex to the green vortex with a double bond,
creating a vertex with two legs emanating from the green vortex and a
single leg from the red vortex; the latter is simply the green vortex
itself.

The numerically found vortex stick network in this model is shown in
fig.~\ref{fig:nonminimal43} and the single and double bonds are
clearly visible on the energy density part (i.e.~upper-left part) of
the figure.
Unfortunately, the end caps possessing triple bonds cannot be seen on
the figure as the triple bonds are too short and weak in the energy
density plots to be seen.
The specific vortex stick network depicted in
fig.~\ref{fig:nonminimal43} is composed of four deformed minimal
vortex sticks.

\begin{figure}[!htp]
  \begin{center}
    \mbox{\includegraphics[scale=\figsize]{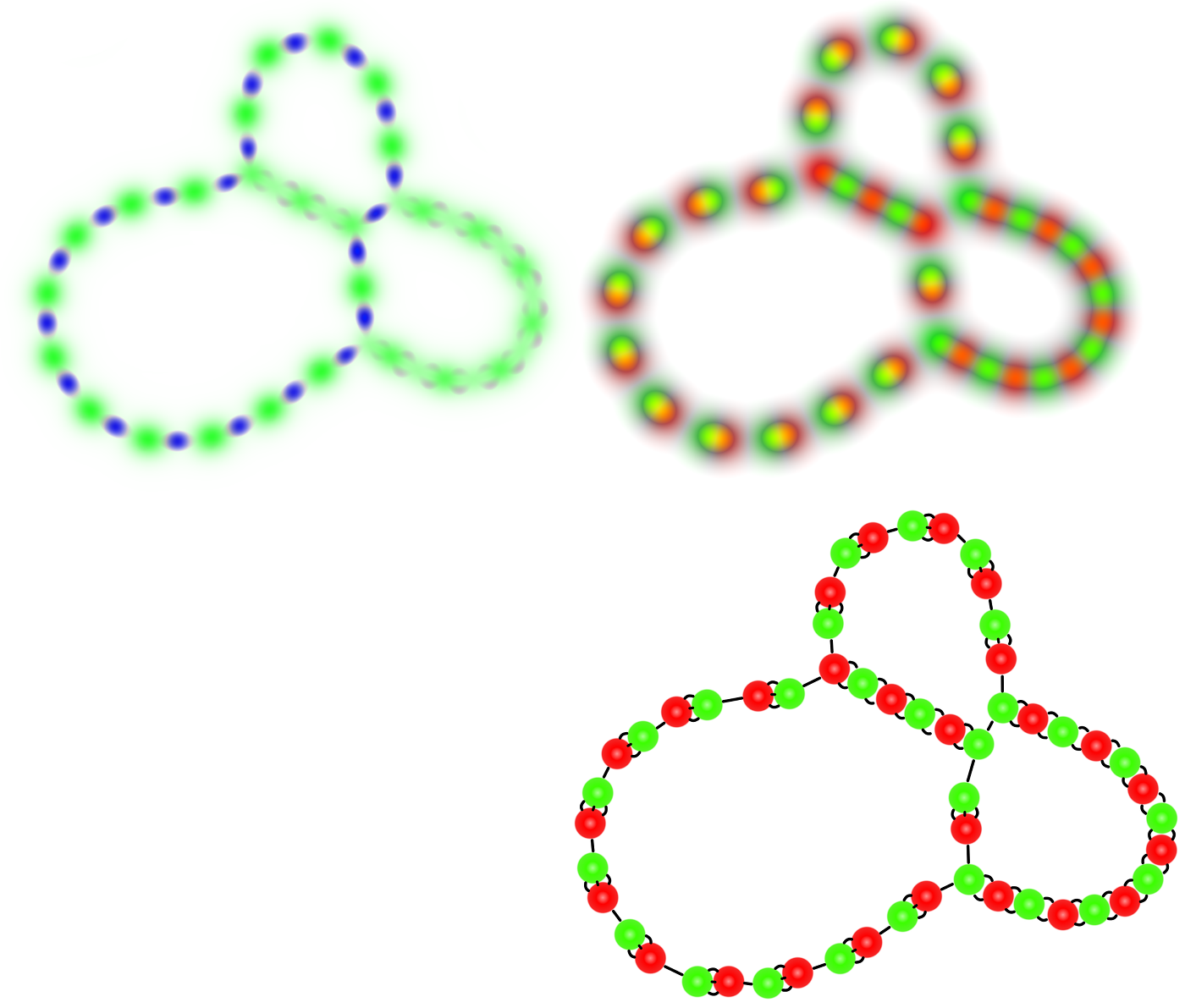}}
    \caption{$(Q_1,Q_2)=(4,4)$ nonminimal vortex ring network. 
    The figure contains 3 elements from left to right and then down:
    an energy density plot, a field density plot and a sketch 
    of how the red and green vortices are interconnected. The first
    two plots are made using the coloring scheme of
    fig.~\ref{fig:color_scheme}.}
    \label{fig:nonminimal44}
  \end{center}
\end{figure}

The final example of nonminimal local vortex solutions is in the
$Q_1=Q_2=4$ model, whose minimal local vortex is simply a ``normal''
vortex composed of a red and green vortex interconnected by a
quadruple bond (i.e.~four legs), see
fig.~\ref{fig:minimal_normal}(d). 
This quadruple bond can be split and thus the minimal vortex can be
turned into components for building a vortex ring network.

The two basic building blocks are two types of links.
The first type of link has a triple bond between a red and a green
vortex, leaving a free leg on both the vortices.
The other type of link possesses only a double bond, providing them
with two free legs on either side.
The second type of link can thus be connected to two links of the
first type.
In principle, vertices do also exist, but are not contained in the
numerical solution displayed in fig.~\ref{fig:nonminimal44}. 
Notice that by connecting two chains of the first type to on chain of
the second type, the two chains of the first type necessarily have the
same vortex species connecting the chain of the second type, and hence
cannot be interconnected.
Therefore, the chains of links are directed graphs and an even number
of insertions of the second type of chain into the first type must
occur.
The numerically found solution of this model in
fig.~\ref{fig:nonminimal44} contains three chains of the first type
and two chains of the second type, thus providing four insertions of
chains of the second type into the ring of the first type.
The single bonds linking the links of the first type and the double
bonds in the chains of the second type are clearly visible in the energy
density part of the figure.
The triple bond possessed by the links of the first type are still not
visible, but are illustrated in the sketch of the figure.

\subsection{Varying the VEVs}\label{sec:varying_vevs}

So far we have worked in a setting where $v_1=v_2=1$, which with
$\lambda_1=\lambda_2$ provides a $\mathbb{Z}_2$ flavor symmetry
between the vortices of the red and green species. 
In this section, we will illustrate how the vortex solutions change by
varying the ratio of the VEVs, $v_1/v_2$.
This could be done in various ways, but we choose to increase either
of the two VEVs, $v_{1,2}$, because that would shrink the vortex of
the given species (as opposed to lowering the VEV that would increase
the vortex).
This is due to the vortex length scales being
\beq
\ell_{\phi_f} = m_{\phi_f}^{-1}
= \frac{1}{\sqrt{2}\lambda_f v_f}, \qquad\textrm{($f$ not summed over)}
\eeq
for the scalar fields and
\beq
\ell_\gamma = m_\gamma^{-1}
= \frac{1}{e\sqrt{Q_1^2v_1^2 + Q_2^2v_2^2}},
\eeq
for the gauge field.
This choice is simply because we use the solutions displayed in the
previous section as seeds for the calculations and increasing the
vortex sizes could make the solutions run out of space on the finite
lattice used for the simulations. 

\begin{figure}[!htp]
  \begin{center}
    \mbox{\subfloat[$(Q_1,Q_2)=(1,1)$]{\includegraphics[scale=\figsize]{vortex_1_1_out}
        \includegraphics[scale=\figsize]{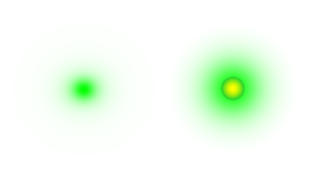}
        \includegraphics[scale=\figsize]{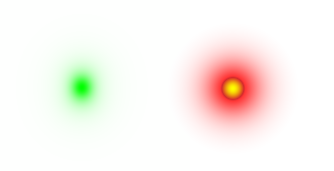}}}
    \caption{$Q_1=Q_2=Q$ minimal ``normal'' vortex with $Q=1$. The
      figure contains 7 elements from left to right: For column 1
      through 3: $v_1/v_2=1$: an energy density plot, a field density
      plot and a sketch of how the red and green vortices are
      interconnected.
    For column 4 and 5: $v_1/v_2=2$: an energy density plot and a
    field density plot.
    For column 6 and 7: $v_1/v_2=1/2$: an energy density plot and a
    field density plot. All columns except column 3 are made using the
    coloring scheme of fig.~\ref{fig:color_scheme}. }
    \label{fig:minimal_normal_vevs}
  \end{center}
\end{figure}

We begin with the minimal ``normal'' vortex with electric charges
$Q_1=Q_2=1$, whose solution is axially symmetric since it is made of a
red and green vortex superposed at exactly the same position.
We can confirm from fig.~\ref{fig:minimal_normal_vevs}, that the
energy density (first column) is axially symmetric and the fields are
coincident (second column), because the field plot is yellow turning
into gray and then into white.
The latter interpolation is exactly the diagonal of both fields
vanishing and while being equal to each other, they increase to their
common VEV.
The third column of fig.~\ref{fig:minimal_normal_vevs} shows the
sketch of the red and the green vortex being interconnected by a
single bond (leg).
The fourth and fifth columns show the same vortex, but for $v_1/v_2=2$
making the red vortex smaller with respect to the equal VEV case shown
in columns one and two.
The result is that the yellow region of the field plot (fourth column)
is small, but surrounded by a green cloud, as the green vortex is
unaltered in size.
We also notice that the energy density plot seems to have shrunk with
respect to the equal VEV case; this is merely because the red vortex
is smaller and hence its energy density is locally higher.
It is still surrounded by a larger and weaker cloud of energy density
coming from the green vortex.
It can be somewhat hard to see, because the energy density plot is
normalized so that the green color is the maximum local energy
density. 
Finally, we increase the second VEV instead of the first, so that
$v_1/v_2=1/2$, which is equivalent to swapping the red and the green
vortex in the configuration.
For this particular case, it is a triviality, and the result can be
seen in columns six and seven of fig.~\ref{fig:minimal_normal_vevs}.
In particular we can see that the small yellow region is now
surrounded by a red cloud instead of a green cloud in the seventh
column of the figure.
For the remaining figures (except fig.~\ref{fig:nonminimal22}) in this
section, $Q_1\neq Q_2$ and hence the sixth and seventh columns of the
figure will be nontrivial compared with the fourth and fifth columns. 

\begin{figure}[!htp]
  \begin{center}
    \mbox{\subfloat[$(Q_1,Q_2)=(2,1)$]{\includegraphics[scale=\figsize]{vortex_2_1_out}
        \includegraphics[scale=\figsize]{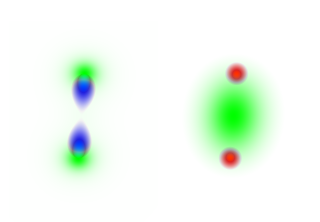}
        \includegraphics[scale=\figsize]{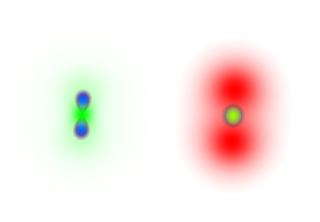}}}
    \mbox{\subfloat[$(Q_1,Q_2)=(3,1)$]{\includegraphics[scale=\figsize]{vortex_3_1_out}
        \includegraphics[scale=\figsize]{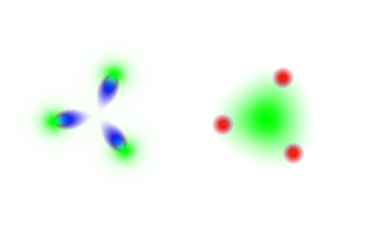}
        \includegraphics[scale=\figsize]{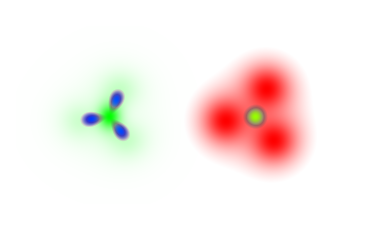}}}
    \mbox{\subfloat[$(Q_1,Q_2)=(4,1)$]{\includegraphics[scale=\figsize]{vortex_4_1_out}
        \includegraphics[scale=\figsize]{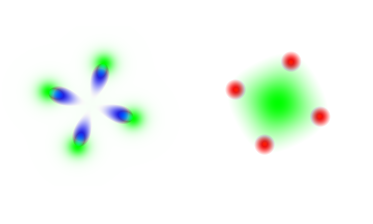}
        \includegraphics[scale=\figsize]{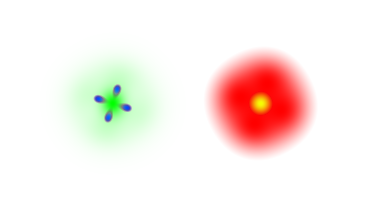}}}
    \caption{$(Q_1,Q_2)=(Q,1)$ minimal vortex ``flowers'', with
      $Q=2,3,4$. Each subfigure contains 7 elements from left to
      right: For column 1 
      through 3: $v_1/v_2=1$: an energy density plot, a field density
      plot and a sketch of how the red and green vortices are
      interconnected.
    For column 4 and 5: $v_1/v_2=2$: an energy density plot and a
    field density plot.
    For column 6 and 7: $v_1/v_2=1/2$: an energy density plot and a
    field density plot. All columns except column 3 are made using the
    coloring scheme of fig.~\ref{fig:color_scheme}. }
    \label{fig:minimal_flowers_vevs}
  \end{center}
\end{figure}

We now turn to varying the VEVs of the minimal vortex ``flowers'', see
fig.~\ref{fig:minimal_flowers_vevs}.
As was the case of the previous figure and for the remaining figures
of this section, the columns show the energy density with $v_1/v_2=1$,
the field densities with $v_1/v_2=1$, the sketch of the vortices and
legs, the energy density with $v_1/v_2=2$, the field densities with
$v_1/v_2=2$, the energy density with $v_1/v_2=1/2$ and finally the
field densities with $v_1/v_2=1/2$.

Reiterating the properties of the vortex flowers for equal VEVs
($v_1/v_2=1$) in fig.~\ref{fig:minimal_flowers_vevs}, we can see all
vortex components in the energy density plots (first column) and the
single bonds are clearly visible in blue as the contribution from the
generalized Josephson term.
From the field densities plot (second column), we can see that the
vortices of both species have the same size.

We now turn to the larger VEV case for the red vortex (i.e.~$v_1/v_2=2$),
see the fourth and fifth columns in
fig.~\ref{fig:minimal_flowers_vevs}.
First we notice that the energy density of the red vortices have
shrunk and are locally so high that the energy density of the green
vortex is barely visible.
The single bonds are still visible, but have shrunk a bit due to the
smaller size of the red vortices.
The reason for the single bond length still being comparatively long
is that the green vortex as the receptacle retains the same size as in
the equal VEV case, thus setting the length scale between the small
red vortices, viz.~the petals.

Turning now to the larger VEV case for the green vortex
(i.e.~$v_1/v_2=1/2$), see the sixth and seventh columns in
fig.~\ref{fig:minimal_flowers_vevs}, we see a similar trend, but since
it is now the receptacle that has shrunk and not the petals, the
solutions are somewhat different in nature.
First we notice that although the green vortex -- the receptacle has
shrunk significantly -- the red vortices are still visible in the
energy density plots, see the sixth column of the figure.
Second, we notice that since the petals are forced to be closer to the
receptacle and to one another, the single bonds binding the petals
have been shortened significantly -- in opposition to the $v_1/v_2=2$
case.
Because the petals -- the red vortices -- are now closer to the
receptacle and to one another, the $(Q_1,Q_2)=(4,1)$ vortex flower has
almost lost its clear discrete $\mathbb{Z}_{Q_1}$ rotational symmetry
(fig.~\ref{fig:minimal_flowers_vevs}(c)), which however is still
visible in the $(Q_1,Q_2)=(2,1)$ and $(Q_1,Q_2)=(3,1)$ solutions
(fig.~\ref{fig:minimal_flowers_vevs}(a,b)). 

\begin{figure}[!htp]
  \begin{center}
    \mbox{\subfloat[$(Q_1,Q_2)=(3,2)$]{\includegraphics[scale=\figsize]{vortex_3_2_a_out}
        \includegraphics[scale=\figsize]{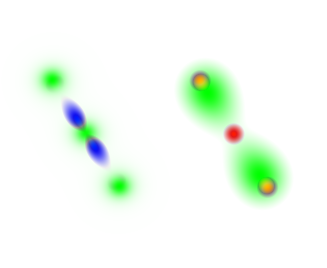}
        \includegraphics[scale=\figsize]{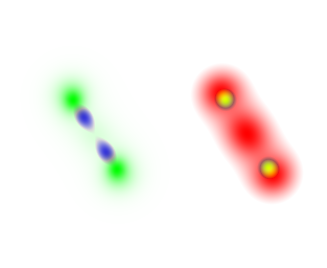}}}
    \caption{$(Q_1,Q_2)=(Q+1,Q)$ vortex ``stick'', with $Q=2$. The
      figure contains 7 elements from left to right: For column 1
      through 3: $v_1/v_2=1$: an energy density plot, a field density
      plot and a sketch of how the red and green vortices are
      interconnected.
    For column 4 and 5: $v_1/v_2=2$: an energy density plot and a
    field density plot.
    For column 6 and 7: $v_1/v_2=1/2$: an energy density plot and a
    field density plot. All columns except column 3 are made using the
    coloring scheme of fig.~\ref{fig:color_scheme}. }
    \label{fig:minimal_sticks_vevs}
  \end{center}
\end{figure}

\begin{figure}[!htp]
  \begin{center}
    \mbox{\subfloat[$(Q_1,Q_2)=(4,2)$]{\includegraphics[scale=\figsize]{vortex_4_2_out}
        \includegraphics[scale=\figsize]{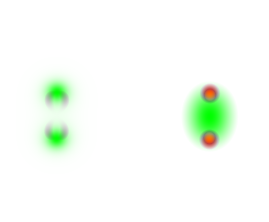}
        \includegraphics[scale=\figsize]{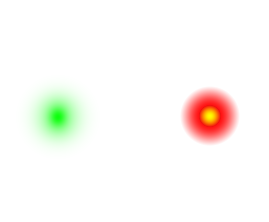}}}
    \caption{$(Q_1,Q_2)=(2Q,Q)$ minimal vortex ``pill'', with $Q=2$. The
      figure contains 7 elements from left to right: For column 1
      through 3: $v_1/v_2=1$: an energy density plot, a field density
      plot and a sketch of how the red and green vortices are
      interconnected.
    For column 4 and 5: $v_1/v_2=2$: an energy density plot and a
    field density plot.
    For column 6 and 7: $v_1/v_2=1/2$: an energy density plot and a
    field density plot. All columns except column 3 are made using the
    coloring scheme of fig.~\ref{fig:color_scheme}. }
    \label{fig:minimal_pills_vevs}
  \end{center}
\end{figure}

The last examples that we will display are the $(Q_1,Q_2)=(3,2)$
vortex ``stick'' and the $(Q_1,Q_2)=(4,2)$ vortex ``pill'', both
possessing double bonds and the former also single bonds, see
figs.~\ref{fig:minimal_sticks_vevs} and \ref{fig:minimal_pills_vevs}, 
respectively.

Starting with the $(Q_1,Q_2)=(3,2)$ vortex ``stick'' of
fig.~\ref{fig:minimal_sticks_vevs}, we again see that the energy
density is localized around the three small red vortices in the fourth
column of the figure, corresponding to the case of $v_1/v_2=2$;
furthermore, the double bonds are tightly connected to the exterior
red vortices and are visible in blue/gray representing the density of
the generalized Josephson term.
The single bonds, on the other hand, are not moved much and are strung
over the length scale of the bond that is roughly set by the size of
the green vortices, which are still being of normal size.
In the other regime, i.e.~$v_1/v_2=1/2$, the green vortices shrink and
the red vortices are moved closer to one another, making the vortex
``stick'' a cloud of red vortex density with small concentrated green
vortices chaining up the cloud, see the sixth and seventh columns of
fig.~\ref{fig:minimal_sticks_vevs}. 
In this case, the double bonds are barely visible, but the single
bonds still are -- although they have shrunk a bit too with respect to
the $v_1/v_2=2$ case.

Finally, we come to the $(Q_1,Q_2)=(4,2)$ vortex ``pill'', which
possesses only double bonds and is shown in
fig.~\ref{fig:minimal_pills_vevs}.
In the fourth and fifth columns, which correspond to the case of
$v_1/v_2=2$, the red vortices are small satellites of the larger green
vortex ``cloud'' and the double bonds are clearly visible in the
fourth column of the figure.
Somewhat surprisingly, the opposite case, $v_1/v_2=1/2$, which only
shrinks the center green vortex of the ``pill'', shows a different
behavior as the red vortices have collapsed into an indistinguishable
cloud with a small green vortex in the center (shown with yellow color
due to the overlap of vortex densities), see the seventh column of
fig.~\ref{fig:minimal_pills_vevs}.

\subsection{Global vortices}\label{sec:global}

In this section, we will consider an example of a global vortex.
This is easily achieved and in fact there is a double infinity of
global vortex solutions.
The criteria for the global vortices is simply
\beq
n_{\rm global} = Q_2 n_1 - Q_1 n_2 \neq 0.
\eeq

For concreteness, we consider the model with $(Q_1,Q_2)=(1,4)$ and
place a single vortex in the first complex scalar field, i.e.~$n_1=1$
and $n_2=0$, yielding
\beq
n_{\rm global} = 4, \qquad n_{\rm local} = 4,
\eeq
and the total winding number
\beq
k = \frac{v_1^2}{v_1^2 + 16v_2^2}.
\eeq
For simplicity, we will also consider only the equal VEV case,
i.e.~$v_1=v_2$, for which the winding number reduces to
\beq
k = \frac{1}{17}.
\eeq

\begin{figure}[!htp]
  \begin{center}
    \mbox{\subfloat[]{\includegraphics[width=0.49\linewidth]{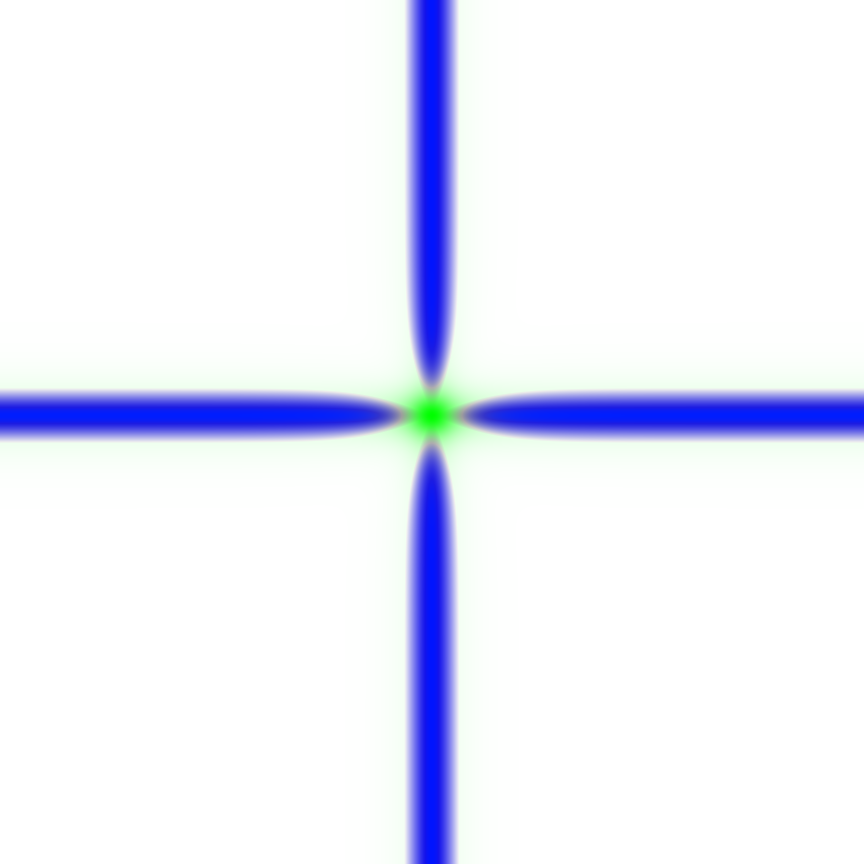}}
      \subfloat[]{\includegraphics[width=0.49\linewidth]{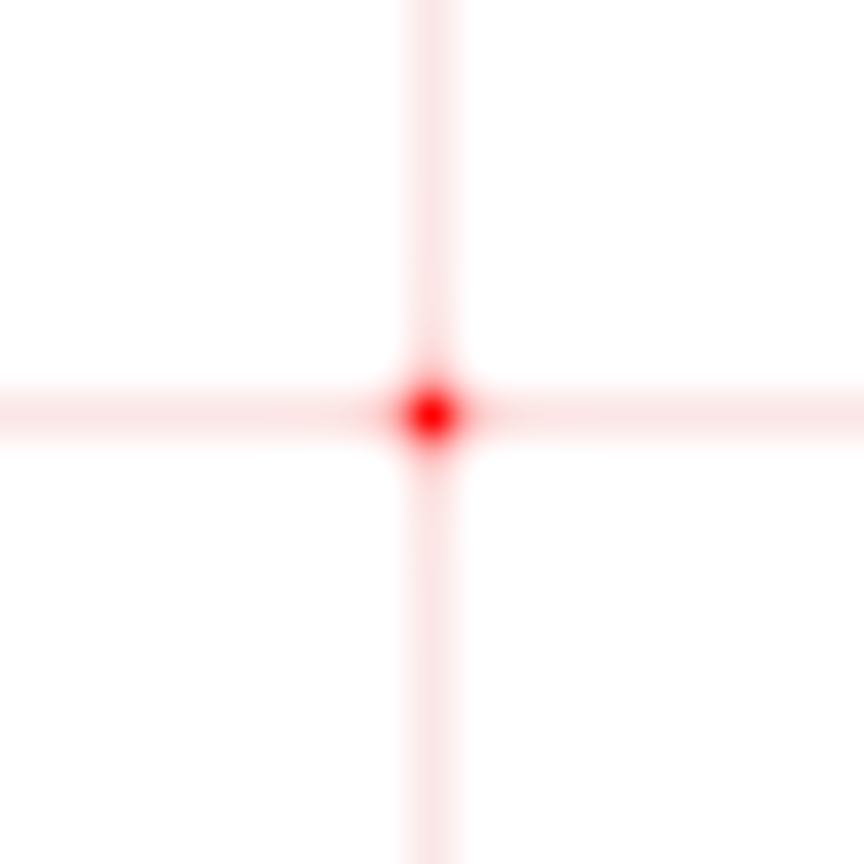}}}
    \mbox{\subfloat[]{\includegraphics[width=0.49\linewidth]{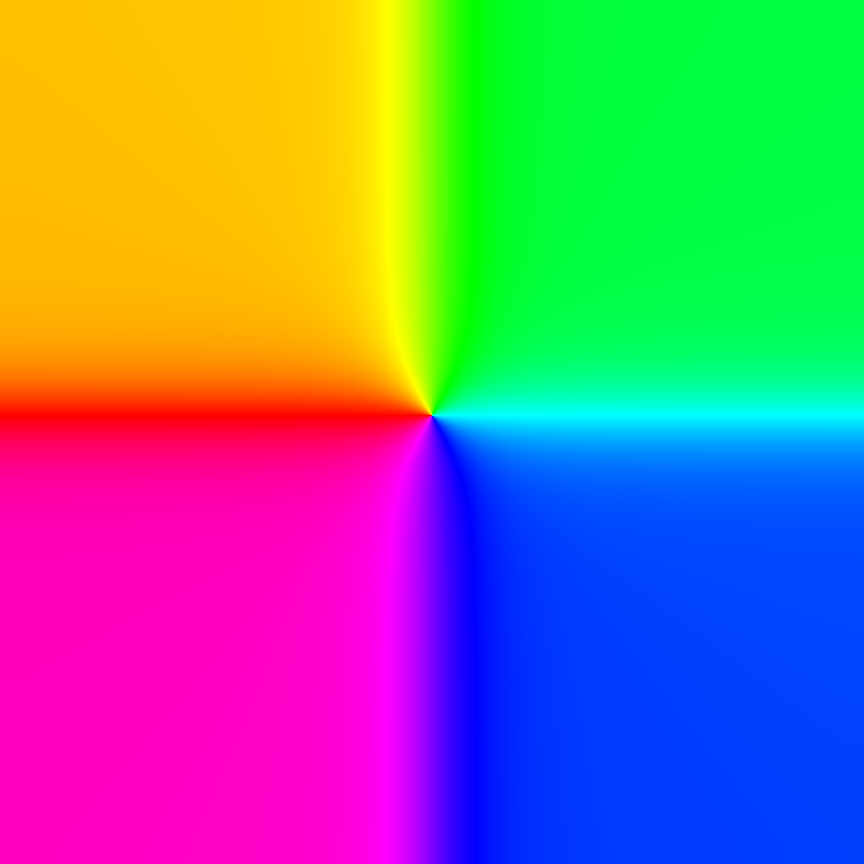}}}
    \caption{$(Q_1,Q_2)=(1,4)$ \emph{global} vortex with four angular
      domain walls: (a) energy density (green) and density of the
      generalized Josephson term (blue) using the color scheme of
      fig.~\ref{fig:color_scheme}(b); (b) field density using the
      color scheme of fig.~\ref{fig:color_scheme}(a); (c)
      $\arg(\phi_1)$ mapped to the color wheel (hue). }
    \label{fig:global14}
  \end{center}
\end{figure}

Fig.~\ref{fig:global14} shows a numerical calculation of the model
with $(Q_1,Q_2)=(1,4)$ with a single vortex in the first scalar field,
namely $n_1=1$ and $n_2=0$.
Fig.~\ref{fig:global14}(a) shows the energy density plot depicting the
total energy density with green and the density of the generalized
Josephson term with blue.
Since the vortex is \emph{global} as opposed to local, there are
$Q_2=4$ domain walls emanating from the vortex core and extending
towards spatial infinity.
Fig.~\ref{fig:global14}(b) shows the field densities and only the red
vortex field is visible; at the origin, the vortex is clearly seen as
the scalar field has zero norm and hence deviates maximally from its
VEV (from below) -- this is marked with red color as opposed to the
vacuum which is white.
At the center of the domain wall (lines), the complex scalar field
deviates slightly from its VEV and thus four dim red lines emanating
from the vortex core are visible in fig.~\ref{fig:global14}(b). 
This is counter to the assumptions on which the analytic solution for
the angular domain walls \eqref{eq:angularDW} were constructed.
Finally, the phase of the complex scalar field, $\phi_1$, is shown in
fig.~\ref{fig:global14}(c) with the argument of the complex scalar
field mapped to the hue of the color wheel.
With this convention $\arg(\phi_1)=0$ is red, $\arg(\phi_1)=2\pi/3$ is
green, $\arg(\phi_1)=4\pi/3$ is blue, $\arg(\phi_1)=\pi/3$ is yellow,
$\arg(\phi_1)=\pi$ is cyan, and finally $\arg(\phi_1)=5\pi/3$ is
magenta. 

\begin{figure}[!htp]
  \begin{center}
    \includegraphics[width=0.5\linewidth]{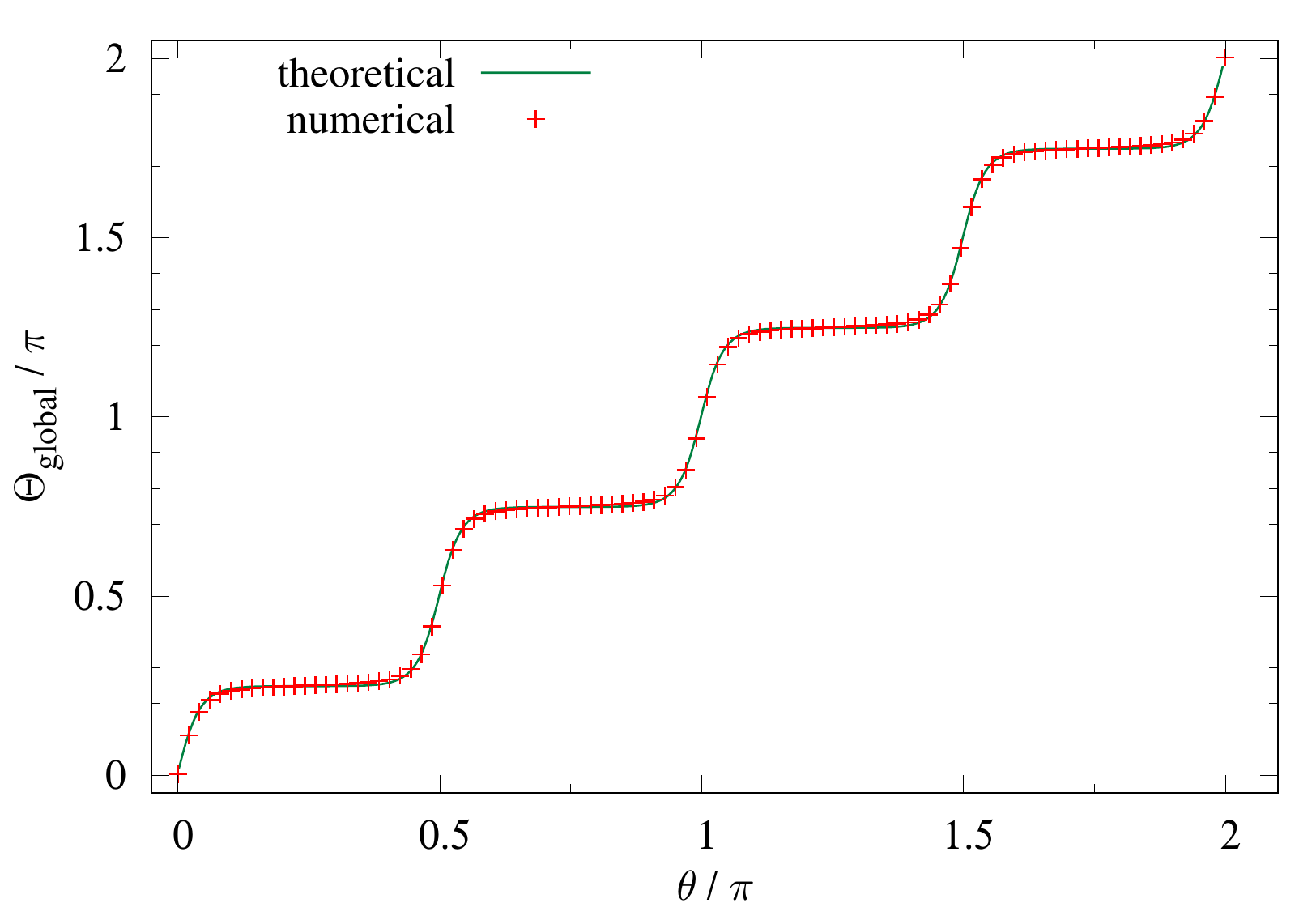}
    \caption{Numerical solution possessing four angular domain
      walls. The theoretical line in green is from
      eq.~\eqref{eq:angularDW} and the red crosses are numerical phase
    data extracted from fig.~\ref{fig:global14}. The data are
    extracted at $r=9$ and here $\kappa=\sqrt{0.1(1+4^2)}\simeq 1.30384$
    (see eq.~\eqref{eq:zeta_kappa_def}). }
    \label{fig:adw4numerical}
  \end{center}
\end{figure}

In fig.~\ref{fig:adw4numerical} we show a comparison of the analytical
prediction of the angular domain wall \eqref{eq:angularDW} and the
numerically extracted phase data from fig.~\ref{fig:global14}.
It is observed from fig.~\ref{fig:adw4numerical}, that the analytic
solution works qualitatively very well and there is only a slight
deviation from the numerical phase data near the centers of the domain
wall.
This deviation is caused by the assumption that the modulus of the
complex scalar field, $\phi_1$, remains at its vacuum value
$|\phi_1|=v_1$, which is only approximately true, see
fig.~\ref{fig:global14}(b).

\section{Conclusion and discussion}\label{sec:conclusion}

In this paper, we have studied a two-component Abelian-Higgs model with
new cross interaction between the two complex scalar fields, which is
inspired by the seminal Josephson term, but generalized to the case of
the complex scalar fields carrying different electric charges.
The new term is manifestly gauge invariant, but does not possess the
physical interpretation of the interaction responsible for the
Josephson effect, where quantum tunneling carries a Cooper pair across
a thin layer to another superconductor. 

After introducing the model and discussing the symmetries and the
vacuum structure, we studied the vortices with and without the
generalized Josephson term for generic charges, $(Q_1,Q_2)$.
Our first result is that the total winding number $k$ (associated with
the winding compensated by the gauge field $A_\theta$) is a rational
fraction when $n_{\rm global}=0$, but arbitrarily fractional when
$n_{\rm global}\neq 0$.
Thereafter, we found the analytic angular domain wall solutions that
exist in the generalized Josephson phase, which we denote the global
vortex phase variable.

We considered also the conditions for having vortices with finite energy
and found that for $\gamma=0$, corresponding to having turned off the
generalized Josephson term, the total energy in the plane is
logarithmically divergent with the cut-off scale (the largest length
in the system) unless $n_{\rm global}=0$, whereas for $\gamma\neq 0$
-- i.e.~with the generalized Josephson term taken into account -- the
total energy is linearly divergent with the cut-off scale, again unless
$n_{\rm global}=0$.
The conclusion is thus that the only finite-energy vortices in the
system, are local vortices (i.e.~with
$n_{\rm global}=Q_2n_1-Q_1n_2=0$) -- independently of whether the
generalized Josephson term is turned on.
The linearly divergent energy, of course, corresponds just to the fact
that there are angular domain walls emanating from the global vortices
and they tend to infinity or to the size of the superconducting
system.

The numerical results of the paper comprise a full classification of
the minimal local vortices, which have the smallest energies among
their respective classes of electric charges.
The classes of minimal local vortices contain a normal vortex, a
vortex flower, a vortex stick, etc.
The nonminimal local vortices exist in an infinite multitude and we
have only provided a few interesting examples.
Generally the nonminimal local vortex configurations appear in two
categories: ring networks and stick networks -- and both can be
infinitely extended by trivial compositions.

Our final result is a comparison of the analytic angular domain wall
solution with those possessed in a numerically found global vortex.
The agreement is very good, even though the assumption of the complex
scalar field to be at its VEV turns out not quite to hold true.

One improvement of the model would be the possibility to admit larger
charges.
If the charges have a common factor, a simple variant of the
generalized Josephson term could be:
\beq
-\gamma\left(\left(\phi_1^{Q_2}\bar{\phi}_2^{Q_1}\right)^{1/\gcd(Q_1,Q_2)}
+ \left(\bar{\phi}_1^{Q_2}\phi_2^{Q_1}\right)^{1/\gcd(Q_1,Q_2)}\right),
\eeq
which nevertheless does not ameliorate the situation of a large charge
\emph{ratio}.
Obviously, the problem of having a power less than one of either of
the fields in the generalized Josephson term leads to an equation of
motion with a point singularity at the vortex positions.

The immediate generalization of our work is to consider $N$-component
superconductors or $N$ flavors of complex scalar fields
\cite{Babaev:2002ck}.
For $N=3$ flavors with equal charge, the minimal local vortex takes
the form of a Y-shaped junction of three fractional vortices
\cite{Nitta:2010yf}. 
Global analogues of BECs with three or more components allow for 
molecules made of $N$ vortex species \cite{Eto:2012rc,Eto:2013spa}.  
It would be interesting to study the details of the solutions for such
multi-component systems with $N\geq 3$ and generic electric charges.

Conventional metallic superconductors with the order parameter (scalar
field) with electric charge $2e$ exhibit a ${\mathbb Z}_2$ topological
order, due to nontrivial linking of a Wilson loop and a vortex,
associated with the spontaneous symmetry breaking of a ${\mathbb Z}_2$
one-form symmetry \cite{Hansson:2004wca}. 
It is an interesting question whether our model with generic charges
exhibits a similar type of topological order.

The more interesting development of this model would be to find a
situation in condensed matter physics or other areas of physics, where
the newly introduced term -- the generalized Josephson term -- is
realized as a physical effect.

\subsection*{Acknowledgments}

S.~B.~G.~thanks Keisuke Ohashi and Calum Ross for discussions. 
C.~C.~is supported by JSPS KAKENHI Grant Number 19K14713.
S.~B.~G.~thanks the Outstanding Talent Program of Henan University for
partial support.
The work of S.~B.~G.~is supported by the National Natural Science
Foundation of China (Grant No.~11675223). 
M.~N.~is supported by the Ministry of Education, Culture, Sports,
Science (MEXT)-Supported Program for the Strategic Research Foundation
at Private Universities ``Topological Science'' (Grant No.~S1511006). 
M.~N.~is also supported in part by JSPS KAKENHI Grant Numbers 16H03984
and 18H01217.
M.~N.~is also supported in part by a Grant-in-Aid for Scientific
Research on Innovative Areas ``Topological Materials Science''
(KAKENHI Grant No.~15H05855) from MEXT of Japan.


\begin{thebibliography}{99}

\bibitem{Abrikosov:1956sx} 
  A.~A.~Abrikosov,
  \emph{On the Magnetic properties of superconductors of the second group},
  Sov.\ Phys.\ JETP {\bf 5}, 1174 (1957)
  [Zh.\ Eksp.\ Teor.\ Fiz.\  {\bf 32}, 1442 (1957)].
  %%CITATION = SPHJA,5,1174;%%

\bibitem{Nielsen:1973cs} 
  H.~B.~Nielsen and P.~Olesen,
  \emph{Vortex Line Models for Dual Strings},
  \href{http://dx.doi.org/10.1016/0550-3213(73)90350-7}{Nucl.\ Phys.\ B {\bf 61}, 45 (1973).}
  %%CITATION = doi:10.1016/0550-3213(73)90350-7;%%

\bibitem{Jaffe:1980}
  A.~Jaffe and C.~Taubes,
  \emph{Vortices and Monopoles -- Structure of Static Gauge Theories},
  Progress in Physics, Birkh\"auser (1980).

\bibitem{Babaev:2004rm} 
  E.~Babaev, A.~Sudbo and N.~W.~Ashcroft,
  \emph{A Superconductor to superfluid phase transition in liquid metallic hydrogen},
  \href{http://dx.doi.org/10.1038/nature02910}{Nature {\bf 431}, 666 (2004)}
  [\href{http://arxiv.org/abs/cond-mat/0410408}{cond-mat/0410408}].
  %%CITATION = doi:10.1038/nature02910;%%

\bibitem{Zohar:2012ay} 
  E.~Zohar, J.~I.~Cirac and B.~Reznik,
  \emph{Simulating Compact Quantum Electrodynamics with ultracold atoms: Probing confinement and nonperturbative effects},
  \href{http://dx.doi.org/10.1103/PhysRevLett.109.125302}{Phys.\ Rev.\ Lett.\  {\bf 109}, 125302 (2012)}
  [\href{http://arxiv.org/abs/arXiv:1204.6574}{arXiv:1204.6574 [quant-ph]}].
  %%CITATION = doi:10.1103/PhysRevLett.109.125302;%%

\bibitem{Moon:2012jf} 
  E.~G.~Moon,
  \emph{Skyrmions with quadratic band touching fermions: A way to achieve charge 4e superconductivity},
  \href{http://dx.doi.org/10.1103/PhysRevB.85.245123}{Phys.\ Rev.\ B {\bf 85}, 245123 (2012)}
  [\href{http://arxiv.org/abs/arXiv:1202.5389}{arXiv:1202.5389 [cond-mat.str-el]}].
  %%CITATION = doi:10.1103/PhysRevB.85.245123;%%

\bibitem{Garaud:2014laa} 
  J.~Garaud and E.~Babaev,
  \emph{Topological defects in mixtures of superconducting condensates with different charges},
  \href{http://dx.doi.org/10.1103/PhysRevB.89.214507}{Phys.\ Rev.\ B {\bf 89}, no. 21, 214507 (2014)}
  [\href{http://arxiv.org/abs/arXiv:1403.3373}{arXiv:1403.3373 [cond-mat.supr-con]}].
  %%CITATION = doi:10.1103/PhysRevB.89.214507;%%

\bibitem{Babaev:2002ck} 
  E.~Babaev,
  \emph{Phase diagram of a planar two band superconductor: Condensation of vortices with fractional flux quantum and existence of a nonsuperconducting superfluid state in this system},
  \href{http://dx.doi.org/10.1016/j.nuclphysb.2004.02.021}{Nucl.\ Phys.\ B {\bf 686}, 397 (2004)}
  [\href{http://arxiv.org/abs/cond-mat/0201547}{cond-mat/0201547}].
  %%CITATION = doi:10.1016/j.nuclphysb.2004.02.021;%%

\bibitem{Herland:2010}
  E.~V.~Herland, E.~Babaev and A.~Sudb\o{},
  \emph{Phase transitions in a three dimensional $U(1)\ifmmode\times\else\texttimes\fi{}U(1)$ lattice London superconductor: Metallic superfluid and charge-$4e$ superconducting states},
  \href{http://dx.doi.org/10.1103/PhysRevB.82.134511}{Phys.\ Rev.\ B {\bf 82}, no. 13, 134511 (2010).}
  [\href{http://arxiv.org/abs/arXiv:1006.3311}{arXiv:1006.3311 [cond-mat.supr-con]}].
  
\bibitem{Josephson:1962zz} 
  B.~D.~Josephson,
  \emph{Possible new effects in superconductive tunnelling},
  \href{http://dx.doi.org/10.1016/0031-9163(62)91369-0}{Phys.\ Lett.\  {\bf 1}, 251 (1962).}
  %%CITATION = doi:10.1016/0031-9163(62)91369-0;%%

\bibitem{Josephson:1974uf} 
  B.~D.~Josephson,
  \emph{The discovery of tunnelling supercurrents},
  \href{http://dx.doi.org/10.1103/RevModPhys.46.251}{Rev.\ Mod.\ Phys.\  {\bf 46}, 251 (1974).}
  %%CITATION = doi:10.1103/RevModPhys.46.251;%%

\bibitem{Fulton:1989}
  T.~A.~Fulton, P.~L.~Gammel, D.~J.~Bishop, L.~N.~Dunkleberger and G.~J.~Dolan,
  \emph{Observation of combined Josephson and charging effects in small tunnel junction circuits},
  \href{http://dx.doi.org/10.1103/PhysRevLett.63.1307}{Phys.\ Rev.\ Lett.\ {\bf 63}, 1307--1310 (1989).}

\bibitem{Wendin:2016}
  G.~Wendin,
  \emph{Quantum information processing with superconducting circuits: a review},
  \href{http://dx.doi.org/10.1088/1361-6633/aa7e1a}{Rep.\ Prog.\ Phys.\ {\bf 80}, 106001 (2017)}
  [\href{http://arxiv.org/abs/arXiv:1610.02208}{arXiv:1610.02208 [quant-ph]}].

\bibitem{Matthews:1999}
  M.~R.~Matthews, B.~P.~Anderson, P.~C.~Haljan, D.~S.~Hall, M.~J.~Holland, J.~E.~Williams, C.~E.~Wieman and E.~A.~Cornell,
  \emph{Watching a Superfluid Untwist Itself: Recurrence of Rabi Oscillations in a Bose-Einstein Condensate},
  \href{http://dx.doi.org/10.1103/PhysRevLett.83.3358}{Phys.\ Rev.\ Lett.\ {\bf 83}, 3358 (1999)}
  [\href{http://arxiv.org/abs/cond-mat/9906288}{cond-mat/9906288}].
  
\bibitem{Williams:2000}
  J.~Williams, R.~Walser, J.~Cooper, E.~A.~Cornell and M.~Holland
  \emph{Excitation of a dipole topological state in a strongly coupled two-component Bose-Einstein condensate},
  \href{http://dx.doi.org/10.1103/PhysRevA.61.033612}{Phys.\ Rev.\ A {\bf 61}, 033612 (2000)}
  [\href{http://arxiv.org/abs/cond-mat/9904399}{cond-mat/9904399}].
  
\bibitem{Zibold:2010}
  T.~Zibold, E.~Nicklas, C.~Gross and M.~K.~Oberthaler
  \emph{Classical Bifurcation at the Transition from Rabi to Josephson Dynamics},
  \href{http://dx.doi.org/10.1103/PhysRevLett.105.204101}{Phys.\ Rev.\ Lett.\ {\bf 105}, 204101 (2010)}
  [\href{http://arxiv.org/abs/arXiv:1008.3057}{arXiv:1008.3057 [cond-mat.quant-gas]}].
  
\bibitem{Nitta:2012xq} 
  M.~Nitta,
  \emph{Josephson vortices and the Atiyah-Manton construction},
  \href{http://dx.doi.org/10.1103/PhysRevD.86.125004}{Phys.\ Rev.\ D {\bf 86}, 125004 (2012)}
  [\href{http://www.arxiv.org/abs/arXiv:1207.6958}{arXiv:1207.6958 [hep-th]}].
  %%CITATION = doi:10.1103/PhysRevD.86.125004;%%

\bibitem{Kobayashi:2013ju} 
  M.~Kobayashi and M.~Nitta,
  \emph{Sine-Gordon kinks on a domain wall ring},
  \href{http://dx.doi.org/10.1103/PhysRevD.87.085003}{Phys.\ Rev.\ D {\bf 87}, no. 8, 085003 (2013)}
  [\href{http://www.arxiv.org/abs/arXiv:1302.0989}{arXiv:1302.0989 [hep-th]}].
  %%CITATION = doi:10.1103/PhysRevD.87.085003;%%

\bibitem{Fujimori:2016tmw} 
  T.~Fujimori, H.~Iida and M.~Nitta,
  \emph{Field theoretical model of multilayered Josephson junction and dynamics of Josephson vortices},
  \href{http://dx.doi.org/10.1103/PhysRevB.94.104504}{Phys.\ Rev.\ B {\bf 94}, no. 10, 104504 (2016)}
  [\href{http://www.arxiv.org/abs/arXiv:1604.08103}{arXiv:1604.08103 [cond-mat.supr-con]}].
  %%CITATION = doi:10.1103/PhysRevB.94.104504;%%

\bibitem{Nitta:2015mma} 
  M.~Nitta,
  \emph{Josephson junction of non-Abelian superconductors and non-Abelian Josephson vortices},
  \href{http://dx.doi.org/10.1016/j.nuclphysb.2015.07.027}{Nucl.\ Phys.\ B {\bf 899}, 78 (2015)}
  [\href{http://www.arxiv.org/abs/arXiv:1502.02525}{arXiv:1502.02525 [hep-th]}].
  %%CITATION = doi:10.1016/j.nuclphysb.2015.07.027;%%

\bibitem{Nitta:2015mxa} 
  M.~Nitta,
  \emph{Josephson instantons and Josephson monopoles in a non-Abelian Josephson junction},
  \href{http://dx.doi.org/10.1103/PhysRevD.92.045010}{Phys.\ Rev.\ D {\bf 92}, no. 4, 045010 (2015)}
  [\href{http://www.arxiv.org/abs/arXiv:1503.02060}{arXiv:1503.02060 [hep-th]}].
  %%CITATION = doi:10.1103/PhysRevD.92.045010;%%

\bibitem{Nitta:2014rxa} 
  M.~Nitta,
  \emph{Non-Abelian Sine-Gordon Solitons},
  \href{http://dx.doi.org/10.1016/j.nuclphysb.2015.04.006}{Nucl.\ Phys.\ B {\bf 895}, 288 (2015)}
  [\href{http://www.arxiv.org/abs/arXiv:1412.8276}{arXiv:1412.8276 [hep-th]}].
  %%CITATION = doi:10.1016/j.nuclphysb.2015.04.006;%%

\bibitem{Branco:2011iw} 
  G.~C.~Branco, P.~M.~Ferreira, L.~Lavoura, M.~N.~Rebelo, M.~Sher and J.~P.~Silva,
  \emph{Theory and phenomenology of two-Higgs-doublet models},
  \href{http://dx.doi.org/10.1016/j.physrep.2012.02.002}{Phys.\ Rept.\  {\bf 516}, 1 (2012)}
  [\href{http://www.arxiv.org/abs/arXiv:1106.0034}{arXiv:1106.0034 [hep-ph]}].
  %%CITATION = doi:10.1016/j.physrep.2012.02.002;%%

\bibitem{Eto:2018hhg} 
  M.~Eto, M.~Kurachi and M.~Nitta,
  \emph{Constraints on two Higgs doublet models from domain walls},
  \href{http://dx.doi.org/10.1016/j.physletb.2018.09.002}{Phys.\ Lett.\ B {\bf 785}, 447 (2018)}
  [\href{http://www.arxiv.org/abs/arXiv:1803.04662}{arXiv:1803.04662 [hep-ph]}].
  %%CITATION = doi:10.1016/j.physletb.2018.09.002;%%

\bibitem{Eto:2018tnk} 
  M.~Eto, M.~Kurachi and M.~Nitta,
  \emph{Non-Abelian strings and domain walls in two Higgs doublet models},
  \href{http://dx.doi.org/10.1007/JHEP08(2018)195}{JHEP {\bf 1808}, 195 (2018)}
  [\href{http://www.arxiv.org/abs/arXiv:1805.07015}{arXiv:1805.07015 [hep-ph]}].
  %%CITATION = doi:10.1007/JHEP08(2018)195;%%

\bibitem{Ustinov:1998}
  A.~V.~Ustinov,
  \emph{Solitons in Josephson junctions},
  \href{http://dx.doi.org/10.1016/S0167-2789(98)00131-6}{Physica D {\bf 123}, 315--329 (1998).}

\bibitem{Tanaka:2001a}
  Y.~Tanaka, 
  \emph{Phase Instability in Multi-band Superconductors},
  \href{http://dx.doi.org/10.1143/JPSJ.70.2844}{J.\ Phys.\ Soc.\ Jp.\ {\bf 70}, 2844 (2001)}

\bibitem{Tanaka:2001b}
  Y.~Tanaka,
  \emph{Soliton in Two-Band Superconductor},
  \href{http://dx.doi.org/10.1103/PhysRevLett.88.017002}{Phys.\ Rev.\ Lett.\ {\bf 88}, 017002 (2001)}.
  
\bibitem{Son:2001td} 
  D.~T.~Son and M.~A.~Stephanov,
  \emph{Domain walls in two-component Bose-Einstein condensates},
  \href{http://dx.doi.org/10.1103/PhysRevA.65.063621}{Phys.\ Rev.\ A {\bf 65}, 063621 (2002)}
  [\href{http://www.arxiv.org/abs/cond-mat/0103451}{cond-mat/0103451 [cond-mat.soft]}].
  %%CITATION = doi:10.1103/PhysRevA.65.063621;%%

\bibitem{Babaev:2001hv} 
  E.~Babaev,
  \emph{Vortices carrying an arbitrary fraction of magnetic flux quantum in two gap superconductors},
  \href{http://dx.doi.org/10.1103/PhysRevLett.89.067001}{Phys.\ Rev.\ Lett.\  {\bf 89}, 067001 (2002)}
  [\href{http://www.arxiv.org/abs/cond-mat/0103451}{cond-mat/0111192}].
  %%CITATION = doi:10.1103/PhysRevLett.89.067001;%%

\bibitem{Kasamatsu:2004tvg} 
  K.~Kasamatsu, M.~Tsubota and M.~Ueda,
  \emph{Vortex molecules in coherently coupled two-component Bose-Einstein condensates},
  \href{http://dx.doi.org/10.1103/PhysRevLett.93.250406}{Phys.\ Rev.\ Lett.\  {\bf 93}, no. 25, 250406 (2004)}
  [\href{http://www.arxiv.org/abs/cond-mat/0406150}{cond-mat/0406150 [cond-mat.mes-hall]}].
  %%CITATION = doi:10.1103/PhysRevLett.93.250406;%%
  
\bibitem{Goryo:2007}
  J.~Goryo, S.~Soma and H.~Matsukawa,
  \emph{Deconfinement of vortices with continuously variable fractions of the unit flux quanta in two-gap superconductors},
  \href{http://dx.doi.org/10.1209/0295-5075/80/17002}{Eur.\ Phys.\ Lett.\ {\bf 80}, 17002 (2007)}
  [\href{http://arxiv.org/abs/cond-mat/0608015}{cond-mat/0608015}].
 
\bibitem{Tylutki:2016mgy} 
  M.~Tylutki, L.~P.~Pitaevskii, A.~Recati and S.~Stringari,
  \emph{Confinement and precession of vortex pairs in coherently coupled Bose-Einstein condensates},
  \href{http://dx.doi.org/10.1103/PhysRevA.93.043623}{Phys.\ Rev.\ A {\bf 93}, no. 4, 043623 (2016)}
  [\href{http://www.arxiv.org/abs/arXiv:1601.03695}{arXiv:1601.03695 [cond-mat.quant-gas]}].
  %%CITATION = doi:10.1103/PhysRevA.93.043623;%%

\bibitem{Eto:2017rfr} 
  M.~Eto and M.~Nitta,
  \emph{Confinement of half-quantized vortices in coherently coupled Bose-Einstein condensates: Simulating quark confinement in a QCD-like theory},
  \href{http://dx.doi.org/10.1103/PhysRevA.97.023613}{Phys.\ Rev.\ A {\bf 97}, no. 2, 023613 (2018)}
  [\href{http://www.arxiv.org/abs/arXiv:1702.04892}{arXiv:1702.04892 [cond-mat.quant-gas]}].
  %%CITATION = doi:10.1103/PhysRevA.97.023613;%%

\bibitem{Cipriani:2013nya} 
  M.~Cipriani and M.~Nitta,
  \emph{Crossover between integer and fractional vortex lattices in coherently coupled two-component Bose-Einstein condensates},
  \href{http://dx.doi.org/10.1103/PhysRevLett.111.170401}{Phys.\ Rev.\ Lett.\  {\bf 111}, 170401 (2013)}
  [\href{http://www.arxiv.org/abs/arXiv:1303.2592}{arXiv:1303.2592 [cond-mat.quant-gas]}].
  %%CITATION = doi:10.1103/PhysRevLett.111.170401;%%

\bibitem{Kobayashi:2018ezm} 
  M.~Kobayashi, M.~Eto and M.~Nitta,
  \emph{Berezinskii-Kosterlitz-Thouless Transition of Two-Component Bose Mixtures with Intercomponent Josephson Coupling},
  \href{http://dx.doi.org/10.1103/PhysRevLett.123.075303}{Phys.\ Rev.\ Lett.\  {\bf 123}, no. 7, 075303 (2019)}
  [\href{http://www.arxiv.org/abs/arXiv:1802.08763}{arXiv:1802.08763 [cond-mat.stat-mech]}].
  %%CITATION = doi:10.1103/PhysRevLett.123.075303;%%

\bibitem{Kawasaki:2013ae} 
  M.~Kawasaki and K.~Nakayama,
  \emph{Axions: Theory and Cosmological Role},
  \href{http://dx.doi.org/10.1146/annurev-nucl-102212-170536}{Ann.\ Rev.\ Nucl.\ Part.\ Sci.\  {\bf 63}, 69 (2013)}
  [\href{http://www.arxiv.org/abs/arXiv:1301.1123}{arXiv:1301.1123 [hep-ph]}].
  %%CITATION = doi:10.1146/annurev-nucl-102212-170536;%%

\bibitem{Eto:2013hoa} 
  M.~Eto, Y.~Hirono, M.~Nitta and S.~Yasui,
  \emph{Vortices and Other Topological Solitons in Dense Quark Matter},
  \href{http://dx.doi.org/10.1093/ptep/ptt095}{PTEP {\bf 2014}, no. 1, 012D01 (2014)}
  [\href{http://www.arxiv.org/abs/arXiv:1308.1535}{arXiv:1308.1535 [hep-ph]}].
  %%CITATION = doi:10.1093/ptep/ptt095;%%

\bibitem{Nitta:2010yf} 
  M.~Nitta, M.~Eto, T.~Fujimori and K.~Ohashi,
  \emph{Baryonic Bound State of Vortices in Multicomponent Superconductors},
  \href{http://dx.doi.org/10.1143/JPSJ.81.084711}{J.\ Phys.\ Soc.\ Jap.\  {\bf 81}, 084711 (2012)}
  [\href{http://www.arxiv.org/abs/arXiv:1011.2552}{arXiv:1011.2552 [cond-mat.supr-con]}].
  %%CITATION = doi:10.1143/JPSJ.81.084711;%%

\bibitem{Eto:2012rc} 
  M.~Eto and M.~Nitta,
  \emph{Vortex trimer in three-component Bose-Einstein condensates},
  \href{http://dx.doi.org/10.1103/PhysRevA.85.053645}{Phys.\ Rev.\ A {\bf 85}, 053645 (2012)}
  [\href{http://www.arxiv.org/abs/arXiv:1201.0343}{arXiv:1201.0343 [cond-mat.quant-gas]}].
  %%CITATION = doi:10.1103/PhysRevA.85.053645;%%

\bibitem{Eto:2013spa} 
  M.~Eto and M.~Nitta,
  \emph{Vortex graphs as N-omers and CP(N-1) Skyrmions in N-component Bose-Einstein condensates},
  \href{http://dx.doi.org/10.1209/0295-5075/103/60006}{EPL {\bf 103}, no. 6, 60006 (2013)}
  [\href{http://www.arxiv.org/abs/arXiv:1303.6048}{arXiv:1303.6048 [cond-mat.quant-gas]}].
  %%CITATION = doi:10.1209/0295-5075/103/60006;%%

\bibitem{Hansson:2004wca} 
  T.~H.~Hansson, V.~Oganesyan and S.~L.~Sondhi,
  \emph{Superconductors are topologically ordered},
  \href{http://dx.doi.org/10.1016/j.aop.2004.05.006}{Annals Phys.\  {\bf 313}, no. 2, 497 (2004)}
  [\href{http://www.arxiv.org/abs/cond-mat/0404327}{cond-mat/0404327 [cond-mat.supr-con]}].
  %%CITATION = doi:10.1016/j.aop.2004.05.006;%%
  
  
\end{thebibliography}
\end{document}